%%%%%%%%%%%%%%%%%%%%%%%%%%%%%%%%%%%%%%%%%%%%%%%%%%%%%%%%%%%%%%%%%%%%%
%%%%%%%%%%%%%%%%%%%   PLAIN  TEX  FILE   %%%%%%%%%%%%%%%%%%%%%%%%%%%%
%%%%%%%%%%%%%%%%%%%%%%%%%%%%%%%%%%%%%%%%%%%%%%%%%%%%%%%%%%%%%%%%%%%%%
% fichero hampgt.tex, 14-03-1996, Tresguerres
\magnification=\magstep1
\hsize=13cm
\vsize=20cm
\overfullrule 0pt
\baselineskip=13pt plus1pt minus1pt
%\baselineskip=2\baselineskip
\lineskip=3.5pt plus1pt minus1pt
\lineskiplimit=3.5pt
\parskip=4pt plus1pt minus4pt

% macro for slash
\def\negenspace{\kern-1.1em}

%\def \Behauptung{  
%= \hbox to 0pt{ \kern -10pt \lower 3pt \vbox to 15pt {\hbox     
%{$\scriptstyle ! \> {} $} \vglue 6pt \hbox {} }} }  

%Macro for section, subsection and equation numbers:

\newcount\secno
\secno=0
\newcount\susecno
\newcount\fmno\def\z{\global\advance\fmno by 1 \the\secno.
                       \the\susecno.\the\fmno}
\def\section#1{\global\advance\secno by 1
                \susecno=0 \fmno=0
                \centerline{\bf \the\secno. #1}\par}
\def\subsection#1{\medbreak\global\advance\susecno by 1
                  \fmno=0
       \noindent{\the\secno.\the\susecno. {\it #1}}\noindent}

%Macro for d'Alembertian:

\def\sqr#1#2{{\vcenter{\hrule height.#2pt\hbox{\vrule width.#2pt
height#1pt \kern#1pt \vrule width.#2pt}\hrule height.#2pt}}}

%Macros for footnotes:

\newcount\refno
\refno=1
\def\y{\the\refno}
\def\myfoot#1{\footnote{$^{(\y)}$}{#1}
                 \advance\refno by 1}

\def\astfoot#1{\footnote{$^{(*)}$}{#1}}
\def\twoastfoot#1{\footnote{$^{(**)}$}{#1}}

%Macro for list of references:

%Macro for not equal:
\def\neq{\hbox{$\,$=\kern-6.5pt /$\,$}}

%Macro for = with * on top:

%Macro for boldface omega (\clom):
%AM font:
%\font\fbg=ambi10\def\clom{\hbox{\fbg\char33}}

%CM font (when using IPS):
%\font\fbg=cmmib10\def\clom{\hbox{\fbg\char33}}

%Macro for section, and equation numbers: (Use \sectio)

\newcount\secno
\secno=0
\newcount\fmno\def\z{\global\advance\fmno by 1 \the\secno.
                       \the\fmno}
\def\sectio#1{\medbreak\global\advance\secno by 1
                  \fmno=0
       \noindent{\the\secno. {\it #1}}\noindent}

%Macro for semidirect product

%Pound sterling:= {\it \$\/}50--00.

%To use any special fonts with IPS, include the corresponding
%CM font definition in your macro or TeX file. In most cases this
%coincides with the AM font except for the first letter, e.g. cmr9
%and amr9 (note: use lower case letters). The one major exception is
%the bold symbols font, which is cmmib10 in place of ambi10, etc.

%===================================================================== 
\centerline{\bf HAMILTONIAN POINCAR\'E GAUGE THEORY OF GRAVITATION}
\vskip 1.0cm 
\centerline{by}
\vskip 1.0cm
\centerline{A. L\'opez--Pinto, A. Tiemblo\astfoot{To my wife.} and R.
Tresguerres\twoastfoot{To my parents.}}
\vskip 1.0cm
\centerline{\it {IMAFF, Consejo Superior de Investigaciones Cient\'ificas,}}
\centerline{\it {Serrano 123, Madrid 28006, Spain}}
\vskip 1.5cm
\centerline{ABSTRACT}\bigskip 
We develop a Hamiltonian formalism suitable to be applied to
gauge theories in the presence of Gravitation, and to Gravity
itself when considered as a gauge theory. It is based on a 
nonlinear realization of the Poincar\'e group, taken as the local 
spacetime group of the gravitational gauge theory, with $SO(3)$
as the classification subgroup. The Wigner--like rotation induced 
by the nonlinear approach singularizes out the role of time and
allows to deal with ordinary $SO(3)$ vectors. We apply the
general results to the Einstein--Cartan action. We study the 
constraints and we obtain Einstein's classical equations in the 
extremely simple form of time evolution equations of the
coframe. As a consequence of our approach, we identify the
gauge--theoretical origin of the Ashtekar variables.
\bigskip\bigskip 
\sectio{\bf{Introduction }}\bigskip 
\bigskip 
The gauge approach to Gravitation$^{(1)}$ constitutes a promising 
departing point in the search for Quantum Gravity. Contrarily to
Einstein's geometrical formulation of General Relativity (GR),
the gauge theories of Gravitation provide a dynamical foundation
of the gravitational interaction analogous to the Yang--Mills
description of the remaining forces characteristic of the standard
model. Gauge theories of Gravity are based on the local
realization of a given spacetime group, which gives rise to the
appearance of connections playing the role of gauge fields. In 
virtue of this approach, one is concerned from the beginning
with dynamics rather than with geometry. In fact, the usual 
geometrical language employed in Gravity can be understood as 
deriving from a particular interpretation of the fundamental 
underlying gauge theory, since the resulting dynamics is formaly
identical to the geometry of a certain spacetime, whose
structure depends on the local spacetime group chosen. One could
say that the geometry of the physical world is a consequence of
the gravitational gauge interactions, i.e. of the dynamics. In
particular, the Riemannian geometry of GR is recovered by
imposing suitable constraints on the gauge fields. However, 
from the viewpoint of the gauge approach, not the geometry but
the interaction described by the connections constitutes the 
primary subject of research. 

Following a number of previous proposals$^{(2)}$, we have recently
shown$^{(3)}$ that the natural way to construct gauge theories
based on local spacetime symmetries, mostly when translations are
concerned, is the nonlinear coset approach due to Coleman et 
al.$^{(4)}$, originally introduced to deal with internal groups.
The nonlinear realizations solve among others the fundamental 
problem of defining coframes in terms of gauge fields. The 
{\it vierbeine} manifest themselves as the (nonlinear)
connections of the translations$^{(5)}$. Furthermore, we have also 
proven that, for any spacetime group including the Poincar\'e
group as a subgroup, the (anholonomic) metric tensor can be
fixed once and for ever to be Minkowskian,$^{(5)}$ in such a way 
that it does not play any dynamical role. The gravitatinal force
is carried exclusively by gauge fields. The curvatures appear as 
the corresponding field strengths. 

The present state of the observational data allows to choose
among various local spacetime groups, such as the affine$^{(6)}$ or
the Poincar\'e$^{(7)}$ group, when one has to decide which one will
play the role of the gauge group of Gravitation. The dynamical 
actions constructed in terms of such groups lead to equally
acceptable classical predictions which scarcely differ from
those of ordinary GR$^{(8)}$. Thus, for the shake of simplicity
and without prejudging the final choice, we will deal with the
simplest spacetime group which includes the Poincar\'e group as a 
subgroup, namely with the Poincar\'e group itself. The standard
Poincar\'e Gauge Theory (PGT) constructed on it$^{(7)}$ is a good 
candidate to become the fundamental theory of Gravitation. In
fact, as far as the Einstein--Cartan action without additional
quadratic terms is concerned, the Euler--Lagrange equations
predict the vanishing of the torsion, thus being the field
equations undistinguishable from those of ordinary GR.
Consequently, at the classical level both theories coincide. 
Nevertheless, due to the wider number of original degrees of 
freedom of the gauge theory, the conditions of vanishing torsion
leading to the Riemannian space of GR are no more given {\it a
priori}, but have to be studied as constraints. Anyway, as we
will see below, the structure of our constraints essentially
coincides with that predicted in the non perturbative approaches
of Quantum Gravity. Not only; in fact, the dynamical variables
of our nonlinear description relate in a simple way with those
introduced by Ashtekar$^{(9)}$.

The standard formulation of the PGT is the Lagrangian
one$^{(7)}$. However, Dirac's standard quantization
procedure$^{(10,11)}$ requires to depart from a well defined
Hamiltonian formalism. Most of the difficulties concerning the 
quantization of Gravitation originate in the lack of a rigorous 
Hamiltonian approach to Einstein's classical theory.
Particularly in the context of gauge theories in so far as
Gravity is involved, it seems to be a hard task to singularize
out the role of time without breaking down the underlying
symmetry properties. In the usual linear gauge realization of 
spacetime groups, the zero--indices present in any tensor valued
p--form stand for time components, so that they are covariant
but not invariant under the group action. This implies that one 
can deal without troubles with the Lagrangian 4-dimensional 
formulation, in which the spacetime is considered as a whole.
But as soon as one separates time by means of a standard 
foliation$^{(12,13)}$ in order to achieve a Hamiltonian 
description, the explicit spacetime invariance gets lost. 

This difficulty arises even in the simple case of Maxwell's 
theory in the presence of Gravitation. Its 4--dimensional
Lagrangian version does not present any problems. But this is 
no longer true when we separate the spacetime into spacelike 
hypersurfaces disposed along a time direction. The foliation is 
performed with respect to a certain timelike vector field 
$n:=\,\partial _t -N^A\partial _A\,$, with $t$ as a time
coordinate. In a flat space, $t$ transforms covariantly under
the Lorentz group, so that the electric field strength $E\,$, 
defined from the whole $U(1)$ field strength $F:=dA\,$ as minus 
its normal part, i.e. as $E:= - n\rfloor F\,$, is at least
a well behaved zero--component of a Lorentz tensor. But the
situation changes if we consider general coordinate
transformations. In this case, the electric field strength as
well as $t$ are no more Lorentz objects. This feature is common 
to the normal part of any p--form representing fields of the
theory if the time direction is chosen to coincide with a
coordinate of the underlying spacetime manifold. A possible
solution to this problem consists in taking the anholonomic 
timelike vector $e_0$ instead of an arbitrary 
$n:=\,\partial _t -N^A\partial _A$ as the time direction. But in
any case $E:= - e_0\rfloor F$ remains a single component of a 
Lorentz tensor. Only an invariant time vector such as the proper
time or similar would allow to deal with normal parts of
p--forms originally defined on a 4--dimensional manifold,
behaving as Poincar\'e scalars.

The possibility of defining such a time direction is important
if we want to construct a Hamiltonian approach to Gravity with 
symmetry properties analogous to those of classical mechanics. 
Relativistic Hamiltonians are in principle not invariant but 
covariant under the Lorentz group. This is a consequence of the 
fact that the Lagrangian density decomposes as 
$L=\,\vartheta ^0\wedge L_{\bot}\,$, being the normal part 
$L_{\bot}:=\,e_0\rfloor L$ a Lorentz covariant 3--form, in terms
of which the Hamiltonian is defined. Notwithstanding, it will be
possible to define a Hamiltonian formalism invariant under 
local Poincar\'e transformations. The key is a nonlinear
realization of the Poincar\'e group with its subgroup $SO(3)$
taken as the classification subgroup. Such nonlinear realization
induces a Wigner--like rotation$^{(14)}$ of the coframe which 
decomposes the fourvector--valued tetrad $\vartheta ^\alpha$
into an $SO(3)$ threevector--valued triad $\vartheta ^a$ plus an
$SO(3)$ scalar $\vartheta ^0\,$. The latter defines an invariant
time in terms of which an invariant foliation of spacetime can
be performed. As a result, we will have at our disposal a 
Poincar\'e invariant formalism expressed in terms of ordinary 
$SO(3)$ tensors and connections. When applied to Gravity,
Einstein's equations simplify enormously.

The paper is organized as follows. In section 2 we introduce the
nonlinear realization of the Poincar\'e group giving rise to the
invariant time 1--form $\vartheta ^0\,$, and accordingly we define
the new $SO(3)$ gravitational variables. Section 3 is devoted to
the Poincar\'e invariant foliation of spacetime. In section 4 the
Hamiltonian formalism is developped, and in section 5 it is
applied to Yang--Mills theories in the presence of Gravity.
Then, in section 6, we generalize the Hamiltonian approach
to Gravitation itself. We study the Einstein--Cartan action as a
relevant example, since it leads to the same Lagrangian field
equations as ordinary GR. We calculate the constraints and the 
Hamiltonian evolution equations. In section 7 we show that they
imply the Lagrangian ones, and in section 8 we present them in a
more suitable $SO(3)$ formulation which reveals that the
Hamiltonian equations are more restrictive than the Lagrangian
ones. Finally, the relation to Ashtekar variables is pointed out
in section 9. The Appendices will be useful for the readers
interested in the technical details.
\bigskip\bigskip

\sectio{\bf{Nonlinear gauge approach to the Poincar\'e group}}
\bigskip 
Let us consider the Poincar\'e group $G=P\,$, with the Lorentz 
generators $L_{\alpha\beta}$ and the translational generators 
$P_\alpha\,$ $(\alpha\,,\beta =\,0,...3\,)$, satisfying the
usual commutation relations as given in (B.1). We want to 
develop a gauge theory of this group presenting the features
of a Hamiltonian formalism capable to predict the time evolution
of physical systems locally defined on 3--dimensional spacelike 
hypersurfaces. In order to do it, we realize the Poincar\'e group 
nonlinearly with its subgroup $H=SO(3)$ as the classification 
subgroup, see Appendix A. This choice of the classification
subgroup automatically leads to the decomposition of the 
fourvector--valued coframes of the standard approaches$^{(7)}$ 
into an $SO(3)$ triplet plus an $SO(3)$ singlet respectively 
--in analogy to a Wigner rotation$^{(14)}$--, the singlet 
characterizing the time component of the coframe. Needles to say
that the nonlinear framework guarantees that the resulting
theory posseses the whole local Poincar\'e symmetry, in spite of
the fact that only the classification subgroup $SO(3)$ manifests
itself explicitly. Thus, the invariance of the time component
of the coframe under $SO(3)$ transformations means in fact
Poincar\'e invariance. We follow the general procedure outlined in
Appendix A. 

We fix the anholonomic invariant metric to have the Lorentzian 
signature 
$$o_{\alpha\beta}:=\,diag(-\,+\,+\,+\,)\,,\eqno(\z)$$
and we decompose the Lorentz generators into boosts $K_a$ and
space rotations $S_a\,$, respectively defined as 
$$K_a :=\,2\,L_{a0}\quad\,,\quad S_a 
:=-\epsilon _a{}^{bc} L_{bc}\qquad\qquad (a=\,1\,,2\,,3)
\,.\eqno(\z)$$
Their commutation relations are given in (B.6). The infinitesimal
group elements of the whole Poincar\'e group become parametrized as  
$$g=\,e^{i\,\epsilon^\alpha P_\alpha }
e^{i\,\beta ^{\alpha\beta}L_{\alpha\beta}} 
\approx\,1+i\,\left(\epsilon ^0 P_0 +\epsilon ^a P_a +\xi ^a K_a
+\theta ^a S_a\,\right)\,,\eqno(\z)$$
and those (also infinitesimal) of the classification SO(3) subgroup
are taken to be 
$$h =e^{i\,{\bf{\Theta}} ^a S_a}\approx\,1
+i\,{\bf{\Theta}} ^a S_a \,.\eqno(\z)$$
In order to realize the Poincar\'e group $P$ on the coset space 
$P/SO(3)$, we make use of the general formula (A.2)
which defines the nonlinear group action, choosing in particular
for the cosets the parametrization
$$c =e^{-i\,x ^\alpha P_\alpha}e^{i\,\lambda ^a K_a}\,,\eqno(\z)$$
where $x ^\alpha \,$ and $\lambda ^a $ are the (finite) coset
parameters. Other parametrizations are possible, leading to equivalent
results. 

According to (A.2) {\it cum} (2.3--5), the variation of the 
translational parameters reads
$$\eqalign{\delta x^0=&-\xi ^a x_a-\epsilon ^0\,,\cr 
\delta x ^a =&\,\,\epsilon ^a{}_{bc}\theta ^b x^c -\xi ^a x^0 
-\epsilon ^a\,,\cr }\eqno(\z)$$
playing these parameters the role of coordinates. On the other hand,
we obtain the variations of the boost parameters of (2.5) as 
$$\delta\lambda ^a =\,\epsilon ^a {}_{bc}\theta ^b \lambda ^c 
+\xi ^a |\lambda |\coth |\lambda | +{{\lambda ^a\lambda _b\xi
^b}\over{|\lambda |^2}}\left( 1-|\lambda |\coth |\lambda |\,\right)
\,,\eqno(\z)$$
being
$$|\lambda |:=\,\sqrt{\lambda _1{}^2 +\lambda _2{}^2 +\lambda _3{}^2 }
\,.\eqno(\z)$$
The nonlinear $SO(3)$ parameter ${\bf{\Theta}} ^a$ in (2.4) is
calculated to be 
$${\bf{\Theta}} ^a =\,\theta ^a +\epsilon ^a{}_{bc}
{{\lambda ^b \xi ^c}\over{|\lambda |}}\tanh\left( {{|\lambda |}
\over 2}\right) \,.\eqno(\z)$$
The relevance of (2.9) becomes evident by considering the action (A.3)
of the Poincar\'e group on arbitrary fields $\psi$ of a given
representation space of the $SO(3)$ group. It reads infinitesimally
$$\delta\psi =\,i\, {\bf{\Theta}} ^a\rho\left( S_a\,\right)\psi
\,,\eqno(\z)$$
being ${\bf{\Theta}} ^a $ precisely the nonlinear $SO(3)$
parameter (2.9), and $\rho\left( S_a\,\right)$ an arbitrary representation 
of the $SO(3)$ group. This general transformation formula shows
how the whole Poincar\'e group projects itself throw (2.10) on the
action of the classification subgroup $SO(3)$ on its
representation fields.

Now we arrive at the most important quantities in a gauge
theory, namely the gauge fields. We will define the suitable
connection for the nonlinear gauge realization in two steps. We
first introduce the ordinary linear Poincar\'e connection $\Omega$
in (A.4--6) as 
$$\Omega :=-i\,{\buildrel (T)\over{\Gamma ^\alpha}} P_\alpha 
           -i\,\Omega ^{\alpha\beta}L_{\alpha\beta} 
          =-i\,{\buildrel (T)\over{\Gamma ^0}} P_0 
           -i\,{\buildrel (T)\over{\Gamma ^a}} P_a 
           +i\,{\buildrel (K)\over{\Gamma ^a}} K_a 
           +i\,{\buildrel (S)\over{\Gamma ^a}} S_a \,,\eqno(\z)$$
which includes the translational and the Lorentz contributions 
${\buildrel (T)\over{\Gamma ^\alpha}}$ and $\Omega
^{\alpha\beta}$ respectively, decomposed in an obvious way. All 
components of (2.11) transform under the linear
action of the group as true Poincar\'e connections. In terms of
them, we define the nonlinear connection (A.4) as
$$\Gamma :=\,c^{-1}\left(d\,+\Omega\,\right) c
          =-i\,\vartheta ^\alpha P_\alpha 
           -i\,\Gamma ^{\alpha\beta}L_{\alpha\beta}
          =-i\,\vartheta ^0 P_0 
           -i\,\vartheta ^a P_a 
           +i\,X^a K_a 
           +i\,A^a S_a \,.\eqno(\z)$$
Contrarily to the components of (2.11), those of (2.12) are divided into
two sets with very different transformation properties. On the
one hand, $A^a$ behaves under the Poincar\'e group as an $SO(3)$ 
connection; on the other hand, the remaining components vary as
$SO(3)$ tensors. In fact we find 
$$\eqalign{\delta\vartheta ^0 &=\,0\cr 
\delta\vartheta ^a &=\,\epsilon ^a{}_{bc}\,{\bf{\Theta}} ^b\,
\vartheta ^c \cr 
\delta X^a &=\,\epsilon ^a{}_{bc}\,{\bf{\Theta}} ^b\,X^c \cr 
\delta A^a  &=-D\,{\bf{\Theta}} ^a :=-\left( d\,{\bf{\Theta}} ^a 
+\epsilon ^a{}_{bc}\,A^b\,{\bf{\Theta}} ^c\,\right)\,.\cr}\eqno(\z)$$
In addition, the trivial metric $\delta _{ab}$ is a natural 
$SO(3)$ invariant. We will use it to raise and lower the 
anholonomic indices, compare with the role of the Minkowski
metric in Ref.(3). The translational nonlinear connections 
$\vartheta ^0\,$, $\vartheta ^a$ in (2.12) are to be identified as the
1--form basis geometrically interpretable as the coframe or 
{\it vierbein}$^{(5,6,7)}$, whereas the vector--valued 1--forms 
$X^a \equiv\,\Gamma ^{0a}$ represent the gauge fields associated
to the boosts, and $A^a\equiv\,{1\over2}\,\epsilon ^a{}_{bc}\,
\Gamma ^{bc}$ play the role of ordinary rotational connections. 
The astonishing fact is that, as we have repeatedly pointed out 
before, the time component $\vartheta ^0$ of the coframe is 
invariant under local Poincar\'e transformations. Let us examine
this fact in more detail. 

The explicit form of the nonlinear translational connection
components in (2.12) reads 
$$\eqalign{\vartheta ^0&:=\,\tilde{\vartheta }^0\cosh |\lambda | 
+\tilde{\vartheta }^a \,{{\lambda _a}\over{|\lambda |}}\sinh
|\lambda |\,,\cr
\vartheta ^a&:=\,\tilde{\vartheta }^a +\tilde{\vartheta }^b 
\,{{\lambda _b\lambda ^a}\over{|\lambda |^2}}\left(\cosh 
|\lambda |-1\right) +\tilde{\vartheta }^0\, {{\lambda ^a}
\over{|\lambda |}}\sinh |\lambda |\,,\cr }\eqno(\z)$$
being $\tilde{\vartheta}^\alpha $ a coframe whose components are
defined as  
$$\eqalign{\tilde{\vartheta }^0&:=\,{\buildrel (T)\over{\Gamma ^0}}
+\left( d\,x^0 -{\buildrel (K)\over{\Gamma ^a}}x_a\,\right)\,,\cr
\tilde{\vartheta }^a&:=\,{\buildrel (T)\over{\Gamma ^a}}
+\left( d\,x^a +\epsilon ^a{}_{bc}{\buildrel (S)\over{\Gamma ^b}}x^c
-{\buildrel (K)\over{\Gamma ^a}}x^0\,\right)\,.\cr}\eqno(\z)$$
This coframe transforms as a Lorentz four--vector under local
Poincar\'e transformations, namely as
$$\eqalign{\delta\tilde{\vartheta }^0 &=
-\xi _a\,\tilde{\vartheta }^a\,,\cr 
\delta\tilde{\vartheta }^a &=\,\,\epsilon ^a{}_{bc}\,\theta ^b\,
\tilde{\vartheta }^c -\xi ^a\,\tilde{\vartheta }^0\,.\cr}\eqno(\z)$$
In fact, the coframe (2.15) coincides with the one we would obtain
by choosing the Lorentz group as the classification subgroup of
the Poincar\'e group, compare with Ref.(3). In order to clarify
the relationship between $\vartheta ^\alpha $ and 
$\tilde{\vartheta}^\alpha $ let us consider the following 
4--dimensional representation of the Lorentz generators:
$$\rho\left(S_1\,\right):=
-i\pmatrix{0\hskip0.2cm 0\hskip0.2cm 0\hskip0.4cm 0\cr 
           0\hskip0.2cm 0\hskip0.2cm 0\hskip0.4cm 0\cr 
           0\hskip0.2cm 0\hskip0.2cm 0-1\cr
           0\hskip0.2cm 0\hskip0.2cm 1\hskip0.4cm 0}
\quad
\rho\left(S_2\,\right):=
-i\pmatrix{0\hskip0.4cm 0\hskip0.2cm 0\hskip0.2cm 0\cr 
           0\hskip0.4cm 0\hskip0.2cm 0\hskip0.2cm 1\cr 
           0\hskip0.4cm 0\hskip0.2cm 0\hskip0.2cm 0\cr
           0-1\hskip0.2cm 0\hskip0.2cm 0}
\quad
\rho\left(S_3\,\right):=
-i\pmatrix{0\hskip0.2cm 0\hskip0.4cm 0\hskip0.2cm 0\cr 
           0\hskip0.2cm 0-1\hskip0.2cm 0\cr 
           0\hskip0.2cm 1\hskip0.4cm 0\hskip0.2cm 0\cr
           0\hskip0.2cm 0\hskip0.4cm 0\hskip0.2cm 0}$$

$$\rho\left(K_1\,\right):=
\,i\pmatrix{0\hskip0.2cm 1\hskip0.2cm 0\hskip0.2cm 0\cr 
           1\hskip0.2cm 0\hskip0.2cm 0\hskip0.2cm 0\cr 
           0\hskip0.2cm 0\hskip0.2cm 0\hskip0.2cm 0\cr
           0\hskip0.2cm 0\hskip0.2cm 0\hskip0.2cm 0}
\quad 
\rho\left(K_2\,\right):=
\,i\pmatrix{0\hskip0.2cm 0\hskip0.2cm 1\hskip0.2cm 0\cr 
           0\hskip0.2cm 0\hskip0.2cm 0\hskip0.2cm 0\cr 
           1\hskip0.2cm 0\hskip0.2cm 0\hskip0.2cm 0\cr
           0\hskip0.2cm 0\hskip0.2cm 0\hskip0.2cm 0}
\quad 
\rho\left(K_3\,\right):=
\,i\pmatrix{0\hskip0.2cm 0\hskip0.2cm 0\hskip0.2cm 1\cr 
           0\hskip0.2cm 0\hskip0.2cm 0\hskip0.2cm 0\cr 
           0\hskip0.2cm 0\hskip0.2cm 0\hskip0.2cm 0\cr
           1\hskip0.2cm 0\hskip0.2cm 0\hskip0.2cm 0}\,.\eqno(\z)$$
The variations (2.16) may be rewritten in terms of (2.17) as
$$\delta\,\pmatrix{\tilde{\vartheta }^0\cr \tilde{\vartheta }^a}
=\,i\,\left[ \xi ^a \rho\left( K_a\,\right)
+\theta ^a \rho\left( S_a\,\right)\,\right]\,
\pmatrix{\tilde{\vartheta }^0\cr \tilde{\vartheta }^a}\,.\eqno(\z)$$
Making now use of the algebraic equality
$$\lambda ^a\lambda ^b\lambda ^c\rho\left( K_a\,\right)
\rho\left( K_b\,\right)\rho\left( K_c\,\right)
=-|\lambda |^2\lambda ^a\rho\left( K_a\,\right)\,,\eqno(\z)$$
we calculate
$$\eqalign{&e^{-i\,\lambda ^a \rho\left(K_a\,\right)}=\,
1-i\,{{\lambda ^a}\over{|\lambda |}}\rho\left( K_a\,\right)
\sinh |\lambda | -{{\lambda ^a\lambda ^b}\over{|\lambda |^2}}
\rho\left(K_a\,\right)\rho\left(K_b\,\right)
\left(\cosh |\lambda | -1\right)\cr
&=\pmatrix{1\hskip0.2cm 0\hskip0.2cm 0\hskip0.2cm 0\cr 
         0\hskip0.2cm 1\hskip0.2cm 0\hskip0.2cm 0\cr 
         0\hskip0.2cm 0\hskip0.2cm 1\hskip0.2cm 0\cr
         0\hskip0.2cm 0\hskip0.2cm 0\hskip0.2cm 1}+
\pmatrix{0\hskip0.2cm \lambda_1\hskip0.2cm \lambda_2\hskip0.2cm \lambda_3\cr 
         \lambda_1\hskip0.2cm 0\hskip0.3cm 0\hskip0.3cm 0\cr 
         \lambda_2\hskip0.2cm 0\hskip0.3cm 0\hskip0.3cm 0\cr
         \lambda_3\hskip0.2cm 0\hskip0.3cm 0\hskip0.3cm 0}
{{\sinh |\lambda |}\over{|\lambda |}}+
\pmatrix{|\lambda |^2\hskip0.2cm 0\hskip0.8cm 0\hskip0.8cm 0\cr 
         0\hskip0.4cm \lambda _1^2\hskip0.2cm 
             \lambda _1\lambda _2\hskip0.2cm \lambda _1\lambda _3\cr 
         0\hskip0.2cm \lambda _1\lambda _2\hskip0.2cm 
             \lambda _2^2\hskip0.4cm \lambda _2\lambda _3\cr
         0\hskip0.2cm \lambda _1\lambda _3\hskip0.2cm 
             \lambda _2\lambda _3\hskip0.4cm \lambda _3^2}
{{\left(\cosh |\lambda | -1\right)}\over{|\lambda |^2}}
\,.\cr}\eqno(\z)$$
Thus we can rewrite (2.14) as
$$\pmatrix{\vartheta ^0\cr \vartheta ^a}=
\,e^{-i\,\lambda ^a \rho\left(K_a\,\right)}
\pmatrix{\tilde{\vartheta }^0\cr \tilde{\vartheta }^a}\,.\eqno(\z)$$
The matrix $e^{-i\,\lambda ^a \rho\left(K_a\,\right)}$ performs
a change of basis leading from the Lorentz covector--valued
1--forms in the r.h.s. of (2.21) to the $SO(3)$ quantities in the
l.h.s., whose variations are specified in (2.13). In fact, taking
into account the transformation properties of the coset
parametre $\lambda ^a$ as given by (2.7), it is easy to verify how 
the nonlinear realization splits the four--dimensional
representation into the SO(3) singlet $\vartheta ^0$ plus the
SO(3) triplet $\vartheta ^a$ respectively. 

Finally, let us end this section introducing the field strengths
corresponding to the gauge potentials we have defined. They will
play the central role in the construction of the gauge theory of
Gravitation in section 6. In order to obtain them, let us
commute two covariant differentials as defined in (A.8). We get 
$${\bf D}\wedge {\bf D} =\,d\,\Gamma
+\Gamma\wedge\Gamma =-i\,T^0 P_0 -i\,T^a P_a 
+i\,{\cal B}^a K_a+i\,{\cal R}^a S_a \,,\eqno(\z)$$
with the torsions 
$$\eqalign{T^0 &:=\,d\,\vartheta ^0 +\vartheta _a\wedge X^a \cr
T^a &:=\,D\,\vartheta ^a +\vartheta ^0\wedge X^a\,,\cr }\eqno(\z)$$
the boost curvature 
$${\cal B}^a :=\,D\,X^a\,,\eqno(\z)$$
and the rotational curvature
$${\cal R}^a :=\,F^a
-{1\over2}\,\epsilon ^a{}_{bc}\,X^b\wedge X^c\,,\eqno(\z)$$
being the $SO(3)$ covariant differentials defined as  
$$\eqalign{D\,\vartheta ^a &:=\,d\,\vartheta ^a +\epsilon ^a{}_{bc}
\,A^b\wedge\vartheta ^c \cr
D\,X^a &:=\,d\,X^a +\epsilon ^a{}_{bc}\,
A^b\wedge X^c \,,\cr }\eqno(\z)$$
and the $SO(3)$ field strength as 
$$F^a :=\,d\,A^a +{1\over2}\,\epsilon ^a{}_{bc}
\,A^b\wedge A^c \,.\eqno(\z)$$
Later we will return back to these definitions. But first we have
to introduce a suitable foliation of the spacetime.
\bigskip\bigskip

\sectio{\bf Poincar\'e invariant foliation of spacetime}\bigskip
Taking advantage of the existence of the invariant time
component of the coframe, it becomes possible to perform an
invariant foliation adapted to it. This will enable us to
define 3--dimensional hypersurfaces in which the gravitational 
$SO(3)$ tensors find their natural seat. 

The general requirement for a foliation to be possible is given
by Frobenius' theorem. Here we summarize briefly one of its 
formulations$^{(15)}$. Let us depart from a 4--dimensional
manifold $M$. A 3--dimensional distribution ${\cal D}$ on M is a
choice of a 3--dimensional subspace ${\cal D}(m)$ of $M_m$ for
each point $m$ in $M\,$. The distribution ${\cal D}$ is said to
be smooth if for each $m$ in $M$ there exist a neighborhood $U$
and a basis of 3 vector fields $X_i$ spanning ${\cal D}$ at each
point of $U$. A submanifold of $M$ whose tangent spaces at
each point coincide with the subspaces determined by the distribution 
${\cal D}$ is called an integral manifold of ${\cal D}$. Frobenius'
theorem establishes that the necessary and sufficient condition
for the existence of integral manifolds of ${\cal D}$ throw each
point of $M$ (what we will call a foliation of the
4--dimensional manifold into 3--dimensional hypersurfaces) is
that, whenever $X_i\,$, $X_j$ belong to ${\cal D}$, then the Lie
derivative of these vectors with respect to each other, namely 
$\left[\,X_i\,,X_j\,\right]\,$, also belongs to ${\cal D}$. (In this
case, ${\cal D}$ is said to be involutive or completely
integrable.) 

Having in mind this general result, we will perform a gauge
adapted foliation which takes into account the particular
coframe defined in the previous section {\it via} the nonlinear 
realization of the Poincar\'e group. From the 1--form basis (2.14)
we define its dual vector basis $e_\alpha $ such that 
$e_\alpha\rfloor\vartheta ^\beta =\,\delta _\alpha ^\beta\,$. 
In the 4--dimensional space, the Lie derivative of the basis
vectors with respect to each other may be expressed in terms of
the anholonomiticity objects $d\vartheta ^\alpha$ as
$$\left[ e_\alpha\,,e_\beta\,\right] =\,\left( e_\alpha\rfloor 
e_\beta\rfloor d\,\vartheta ^\gamma\,\right) e_\gamma\,.\eqno(\z)$$
Let us now separate $e_{_0}$ from $e_a$ $(a=\,1\,,2\,,3\,)\,$, since
we are interested in foliating the spacetime into spacelike
hypersurfaces equipped with the $SO(3)$ vector bases $e_a$. 
According to Frobenius' theorem, the necessary and sufficient
condition for such a foliation to be possible is that
$$\left[ e_a\,,e_b\,\right] =\,\left( e_a\rfloor 
e_b\rfloor d\,\vartheta ^c\,\right) e_c\,,\eqno(\z)$$
which is equivalent to impose the condition 
$\left( e_a\rfloor e_b\rfloor d\vartheta ^0\,\right) =\,0\,$, or
$d\vartheta ^0 =\vartheta ^0\wedge
\left( e_{_0}\rfloor d\,\vartheta ^0\,\right)\,$. In other words,
$$\vartheta ^0\wedge d\,\vartheta ^0 =\,0\,.\eqno(\z)$$
Notice that, according to (2.13a), eq.(3.3) is Poincar\'e
invariant. In fact, due to the ocurrence in it of an ordinary
instead of a covariant differential, the condition (3.3)
required by the Frobenius theorem is groupally well defined only
if $\vartheta ^0$ is invariant. Accordingly, (3.3) defines an
invariant foliation. From it follows that the time coframe takes
the form
$$\vartheta ^0 =\,u^0\,d\,\tau\,,\eqno(\z)$$
with $\tau$ as a parametric time.

Let us now see how the dual vector $e_{_0}$ of (3.4) looks like. In 
terms of the vector field $\partial _\tau$ associated to the
time parametre $\tau$, we define
$$u^\alpha :=\,\partial _\tau\rfloor\vartheta ^\alpha\,,\eqno(\z)$$
which represents a sort of (coordinate independent) fourvelocity
due to the similarity between $\tau$ and the proper time. We can
express $\partial _\tau$ in terms of the vector basis $e_\alpha$ as
$$\partial _\tau =\,u^\alpha\,e_\alpha =\,u^0\,e_{_0} 
+u^a\,e_a \,.\eqno(\z)$$
From (3.6) follows that the Poincar\'e invariant vector $e_{_0}$ reads
$$e_{_0} =\,{1\over{u^0}}\partial _\tau 
-{{u^a}\over{u^0}}\,e_a \,.\eqno(\z)$$
We can alternatively rewrite it in terms of the
three--dimensional velocity $v^a :=\,c\,u^a/u^0$ or of
the coordinate velocity $v^A:=e^A{}_a\,v^a$ respectively as 
$$e_{_0} =\,{1\over{u^0}}\partial _\tau -{{v^a}\over{c}}\,e_a 
=\,{1\over{u^0}}\partial _\tau -{{\,\,v^A}\over{c}}\,\partial _A
\,.\eqno(\z)$$
We identify $e_{_0}$ as the invariant timelike vector field along
which the foliation of the spacetime is defined. Observe that
the components of the fourvelocity in (3.7) relate in a simple way
to the usual lapse and shift functions considered in the usual
foliations of spacetime$^{(12,13)}$. The main difference between them
and our approach consists in that we foliate the underlying
manifold taking the gauge properties of the coframe, as they
arise from the nonlinear approach, as a guide. Accordingly, in
the following we only will need to consider the lapse--like
function $u^0$ explicitly, since the shift--like velocity $u^a$
will appear enclosed in the $SO(3)$ triads representing the space
coframe, namely
$$\vartheta ^a =\,u^a\,d\,\tau + e^a{}_A d\,x^A\,,\eqno(\z)$$
compare with (3.4). Obviously
$$e_{_0}\rfloor\vartheta ^0 =\,1\,,\eqno(\z)$$
$$e_{_0}\rfloor\vartheta ^a =\,0\,.\eqno(\z)$$
Any arbitrary p--form $\alpha$ admits a decomposition into a
normal and a tangential part with respect to the invariant
vector field $e_{_0}$ inducing the foliation, namely
$$\alpha =\,^{\bot}\alpha +\underline{\alpha }\,,\eqno(\z)$$
with the normal and tangential parts respectively defined as 
$$^{\bot}\alpha :=\,\vartheta ^0\wedge\alpha _{\bot}
\quad\,,\qquad \alpha _{\bot}:=\,e_{_0}\rfloor\alpha \,,\eqno(\z)$$
and
$$\underline{\alpha }:=\,e_{_0}\rfloor\left(\, 
\vartheta ^0\wedge\alpha\,\right) \,.\eqno(\z)$$
In particular we find 
$$\eqalign{\vartheta ^0 &=\,^{\bot}\vartheta ^0 
=\,u^0\,d\,\tau\cr 
\vartheta ^a &=\,\,\,\,\underline{\vartheta }^a \,,\cr }\eqno(\z)$$
thus showing trivially that $\vartheta ^0$ only presents a normal
contribution, and $\vartheta ^a$ only a tangential one. The
general expressions for the foliation of an arbitrary p--form 
$\alpha $ and its Hodge dual read respectively 
$$\eqalign{\alpha &=\,\vartheta ^0\wedge\alpha _{\bot }
+\underline{\alpha }\cr 
^*\alpha &=\,\left( -1\right) ^p\vartheta ^0
\wedge {}^\#\underline{\alpha }-{}^\#\alpha _{\bot }\,.\cr } 
\eqno(\z)$$
The asterisc $*$ stands for the Hodge dual in four dimensions,
whereas $\#$ represents its three--dimensional restriction, see
Appendix E. 

Let us define the Lie derivative with respect to $e_{_0}\,$,
namely 
$${\it{l}}_{e_{_0}}\alpha :=
\,d\,\left( e_{_0}\rfloor\alpha\,\right) 
+ \left( e_{_0}\rfloor d\,\alpha\,\right) \,.\eqno(\z)$$
In particular, the Lie derivative of the tangential part of a
form reduces to
$${\it{l}}_{e_{_0}}\underline{\alpha}
=\,\left( e_{_0}\rfloor d\,\underline{\alpha }\,\right)\,.\eqno(\z)$$
It represents the time evolution of $\underline{\alpha }\,$. Now
we can decompose the exterior differential of an arbitrary form as 
$$d\,\alpha =\,\vartheta ^0\wedge\left[\,{\it{l}}_{e_{_0}}
\underline{\alpha}-{1\over{u^0}}\,\underline{d}\,
\left( u^0\,\alpha _{\bot }\right)\,\right] +\underline{d}\,
\underline{\alpha}\,.\eqno(\z)$$
Observe that the Lie derivative of the normal part 
$\alpha _{\bot}$ is absent. This is a general feature of gauge
theories. Finally, we have at our disposal all the elements 
necessary to develop a Hamiltonian approach which is well
behaved in the presence of Gravity when considered as the
nonlinear gauge theory of the Poincar\'e group.
\bigskip\bigskip

\sectio{\bf Hamiltonian formalism}\bigskip 
In this section we will outline the deduction of the Hamiltonian
treatment of gauge theories in general, expressed in terms of 
differential forms. We follow essentially the work of
Wallner$^{(13)}$, but we make use of the invariant spacetime
foliation of the previous section. The resulting formalism 
resembles strongly the classical one. However, being the dynamical
variables differential forms, the theory is automatically 
diff--invariant. Furthermore, it makes it possible to take into
account the gravitational effects, see sections 5,6 below.

Let us consider a general gauge theory depending on the gauge 
potential $A\,$. When we consider a particular group, $A$
will posses the corresponding group indices, but we do not need
to specify them at this stage. The Lagrangian density depends on
$A$ and on its exterior differential $dA\,$, i.e.
$$L=\,L\left( A\,,d\,A\,\right)\,.\eqno(\z)$$
Since the gauge potential $A$ is a connection, it is a 1--form 
which, as established in the previous section, can be decomposed
into its normal and tangential parts as 
$$A =\,\vartheta ^0\wedge A_{\bot} +\underline{A}\,.\eqno(\z)$$
The Lagrangian density is a 4--form. Thus, its tangential part
does not exist. It decomposes simply as
$$L =\,\vartheta ^0\wedge L_{\bot}\,.\eqno(\z)$$
In terms of its normal part $L_{\bot}$ we define the momenta
$$^\#\pi ^{^{A_{\bot}}} :={{\partial L_{\bot}}\over
{\partial ({\it{l}}_{e_{_0}} A_{\bot})}}\quad\,,\qquad
^\#\pi ^{^{\underline{A}}} :={{\partial L_{\bot}}\over
{\partial ({\it{l}}_{e_{_0}}\underline{A}\,)}}\,,\eqno(\z)$$
and we construct the Hamiltonian 3--form
$${\cal{H}}:=\,u^0\,\left[\,{\it{l}}_{e_{_0}} 
A_{\bot}{}^\#\pi ^{^{A_{\bot}}} +{\it{l}}_{e_{_0}}\underline{A}
\wedge {}^\#\pi ^{^{\underline{A}}} -L_{\bot}\,\right]\,.\eqno(\z)$$
In Appendix C we present a deduction of the Hamilton equations
based on the comparison of the variations of (4.1) and (4.5)
respectively. We do so in order to show the correctness of the
results we will obtain here proceeding in a shorter and more
general way, see Ref.(11). In particular, Appendix C guarantees
the consistence of the definitions of the Lie derivatives used
in (4.7,8) below.

We begin by reconstructing the Lagrangian density in terms of
the momenta and (4.5), namely
$$L =\,d\,A_{\bot}\wedge{}^\#\pi ^{^{A_{\bot}}} +d\,\underline{A}\wedge
{}^\#\pi ^{^{\underline{A}}} -d\,\tau\wedge {\cal{H}}\,,\eqno(\z)$$
and we deduce the field equations 
$$u^0\,{\it{l}}_{e_{_0}} A_{\bot}
:={{d\,A_{\bot}}\over{d\,\tau}} 
={{\delta {\cal{H}}}\over{\delta\,{}^\#\pi ^{^{A_{\bot}}}}}\quad\,,\qquad
u^0\,{\it{l}}_{e_{_0}}{}^\#\pi ^{^{A_{\bot}}} 
:={{d\,{}^\#\pi ^{^{A_{\bot}}}}\over{d\,\tau}}
=-{{\delta {\cal{H}}}\over{\delta A_{\bot}}}\,,\eqno(\z)$$
$$u^0\,{\it{l}}_{e_{_0}} \underline{A} 
:={{d\,\underline{A}}\over{d\,\tau}}
={{\delta {\cal{H}}}\over{\delta\,{}^\#\pi ^{^{\underline{A}}}}}
\quad\,,\qquad
u^0\,{\it{l}}_{e_{_0}}{}^\#\pi ^{^{\underline{A}}}
:={d\,{{}^\#\pi ^{^{\underline{A}}} }\over{d\,\tau}}
=-{{\delta {\cal{H}}}\over{\delta \underline{A}}}\,.\eqno(\z)$$
On the other hand, the Lie derivative of an arbitrary p--form
defined on the 3--space and being a functional of the dynamical 
variables may be expanded as
$${\it{l}}_{e_{_0}} \omega =\,
{\it{l}}_{e_{_0}} A_{\bot}
\,{{\delta \omega }\over{\delta A_{\bot}}}
+{\it{l}}_{e_{_0}}\underline{A}\wedge {{\delta \omega }
\over{\delta \underline{A}}}
+{{\delta \omega }\over{\delta\,{}^\#\pi ^{^{A_{\bot}}}}}
\,{\it{l}}_{e_{_0}}{}^\#\pi ^{^{A_{\bot}}}   
+{{\delta \omega }\over{\delta\,{}^\#\pi ^{^{\underline{A}}}}}
\wedge {\it{l}}_{e_{_0}}{}^\#\pi ^{^{\underline{A}}}\,.\eqno(\z)$$
We substitute the field equations (4.7,8) into (4.9) and accordingly we
define Poisson brackets representing the time evolution of 
differential forms as
$$\eqalign{u^0\,{\it{l}}_{e_{_0}} \omega 
=\,\left\{\omega\,,{\cal{H}}\,\right\} &:=
{{\delta {\cal{H}}}\over{\delta\,{}^\#\pi ^{^{A_{\bot}}}}}
\,{{\delta \omega}\over{\delta A_{\bot}}}
-{{\delta \omega}\over{\delta\,{}^\#\pi ^{^{A_{\bot}}}}}\,
{{\delta {\cal{H}}}\over{\delta A_{\bot}}}\cr
&\hskip0.2cm+{{\delta {\cal{H}}}\over{\delta\,
{}^\#\pi ^{^{\underline{A}}}}}\wedge {{\delta
\omega}\over{\delta \underline{A}}}
-{{\delta \omega}\over{\delta\,
{}^\#\pi ^{^{\underline{A}}}}}
\wedge {{\delta {\cal{H}}}\over{\delta 
\underline{A}}}\,.\cr }\eqno(\z)$$
From definition (4.10) we obtain the basic properties of the Poisson
brackets. In the first place, they are evidently antisymmetric:
$$\left\{\omega\,,{\cal{H}}\,\right\} 
=-\left\{{\cal{H}}\,,\omega\,\right\}\,.\eqno(\z)$$
Taking into account the chain rule of the Lie derivative, namely
$${\it{l}}_{e_{_0}}\left( \sigma\wedge\omega\,\right) 
=\,{\it{l}}_{e_{_0}}\sigma\wedge\omega  
+\sigma\wedge {\it{l}}_{e_{_0}}\omega \,,\eqno(\z)$$
we get the distributive property
$$\left\{\sigma\wedge\omega\,,{\cal{H}}\,\right\} 
=\,\left\{\sigma\,,{\cal{H}}\,\right\}\wedge\omega  
+\sigma\wedge\left\{\omega\,,{\cal{H}}\,\right\} \,.\eqno(\z)$$
Finally, since the normal part of the identity 
$d\wedge d\,\alpha \equiv\,0$ reads
$${\it{l}}_{e_{_0}} \underline{d}\,\underline{\alpha} 
-{1\over{u^0}}\,\underline{d}\left( u^0\,
{\it{l}}_{e_{_0}} \underline{\alpha }\,\right)\equiv\,0
\,,\eqno(\z)$$
it follows that, being $\omega$ a form defined on the 3--space 
and thus being absent a normal part of it, it holds 
$$\left\{\,\underline{d}\,\omega\,,{\cal{H}}\,\right\}
-\underline{d}\left\{ \omega\,,{\cal{H}}\,\right\}
=\,0\,.\eqno(\z)$$
This completes the formal instrument we need to calculate the
time evolution of any dynamical variable in terms of the 
Hamiltonian 3--form (4.5). 
\bigskip\bigskip 

\sectio{\bf Yang--Mills theories}\bigskip
Instead of directly undertake the study of the Hamiltonian
treatment of Gravitation, let us first consider the example of
simpler gauge theories. This will illuminate the further
developments, which are a straightforward generalization of
what we will establish in this section. In particular, we will
show how to deal with constraints.

Let us consider the Yang--Mills theory of an internal group with
structure constants $f^a{}_{bc}\,$, antisymmetric in the two
last indices. Its field strength 
$$F^a :=\,d\,A^a +{1\over2}\,f^a{}_{bc}\,A^b\wedge A^c\eqno(\z)$$
decomposes into 
$$F^a =\,\vartheta ^0\wedge F_{\bot}^a +\underline{F}^a\,,\eqno(\z)$$
being its normal and tangential parts respectively given by 
$$F_{\bot}^a :=\,{\it{l}}_{e_{_0}}\,\underline{A}^a
-{1\over{u^0}}\,\underline{D}\left(\,u^0\, A_{\bot}^a\,\right)
\quad\,,\qquad 
\underline{F}^a :=\,\underline{d}\,\underline{A}^a 
+{1\over 2}\,f^a{}_{bc}\,\underline{A}^b\wedge\underline{A}^c 
\,.\eqno(\z)$$
According to (3.16b), its Hodge dual decomposes as
$$^*F^a =\,\vartheta ^0\wedge {}^\#\underline{F}^a 
- {}^\#F_{\bot}^a \,.\eqno(\z)$$
In the absence of matter, the Lagrangian density of a standard
Yang--Mills theory is given by
$$L =-{1\over2}\,F^a\wedge {}^*F_a =\,{1\over2}\,\vartheta ^0
\wedge\left( F_{\bot}^a \wedge {}^\#F_{\bot}{}_a 
-\underline{F}^a \wedge {}^\#\underline{F}{}_a \,\right)\,.\eqno(\z)$$ 
From (5.5) we get the normal part
$$L _{\bot} =\,{1\over 2}\,\left( F_{\bot}^a \wedge {}^\#F_{\bot}{}_a 
-\underline{F}^a \wedge {}^\#\underline{F}{}_a \,\right)\,,\eqno(\z)$$ 
and we calculate the momenta as defined in (4.4), namely
$$^\#\pi ^{^{A_{\bot}}}_a 
:=\,{{\partial L_{\bot}}\over{\partial\left( 
{\it{l}}_{e_{_0}} A_{\bot}^a\,\right)}}=\,0\,,\eqno(\z)$$
$$^\#\pi ^{^{\underline{A}}}_a 
:=\,{{\partial L_{\bot}}\over{\partial\left( 
{\it{l}}_{e_{_0}}\underline{A}^a\,\right)}}=\,{}^\#F_{\bot}^a 
\,.\eqno(\z)$$
The first one is a primary constraint. We will come back
to this fact immediately to show how to proceed in this and 
analogous cases. For the present, we calculate the canonical 
Hamiltonian 3--form (4.5) as
$$\eqalign{{\cal{H}}_0 :=&\,\,u^0\left(\,
{\it{l}}_{e_{_0}} A_{\bot}^a\,\,{}^\#\pi ^{^{A_{\bot}}}_a 
+{\it{l}}_{e_{_0}}\underline{A}^a
\wedge\,{}^\#\pi ^{^{\underline{A}}}_a -L_{\bot}\right)\cr 
=&\,{1\over2}\,u^0\left( \pi ^{^{\underline{A}}}{}^a
\wedge {}^\#\pi ^{^{\underline{A}}}_a 
+\underline{F}^a \wedge {}^\#\underline{F}{}_a \,\right)
+\underline{D}\left(\,u^0\, A_{\bot}^a\,\right)\wedge 
{}^\#\pi ^{^{\underline{A}}}_a \,.\cr }\eqno(\z)$$
But the time evolution operator must include information about
the primary constraint (5.7). This is carried out by adding it by
means of a Lagrange multiplier$^{(11)}$. Doing so and integrating 
by parts the last term in (5.9), we get the total Hamiltonian 
$${\cal{H}} :=\,{1\over2}\,u^0\left( \pi ^{^{\underline{A}}}{}^a
\wedge {}^\#\pi ^{^{\underline{A}}}_a 
+\underline{F}^a \wedge {}^\#\underline{F}{}_a \,\right)
-u^0\, A_{\bot}^a\,\underline{D}\,{}^\#\pi ^{^{\underline{A}}}_a 
+\beta ^a\,{}^\#\pi ^{^{A_{\bot}}}_a \,.\eqno(\z)$$ 
Observe that the result is equivalent to having departed from
the definition (4.5) and having replaced the Lie derivative
acompanying the constraint (5.7) by a Lagrange multiplier. We 
retain this recipe to procede analogously in the following.

The function $u^0$ appearing everywhere, see (3.4), carries
information about the underlying geometry and thus about the
gravitational background. Nevertheless, relatively to the
internal group considered here $u^0$ does not play the role of a
dynamical variable. Only $A_{\bot}^a$ and $\underline{A}^a$ play
this role, so that the Poisson brackets (4.10) take the 
form
$$\eqalign{u^0\,{\it{l}}_{e_{_0}}\omega 
=\,\left\{\omega\,,{\cal{H}}\,\right\} &:=
{{\delta {\cal{H}}}\over{\delta\,{}^\#\pi ^{^{A_{\bot}}}_a}}
\,{{\delta \omega}\over{\delta A_{\bot}^a}}
\,\,-\,{{\delta \omega}\over{\delta\,{}^\#\pi ^{^{A_{\bot}}}_a}}
\,{{\delta {\cal{H}}}\over{\delta A_{\bot}^a}}\cr
&\hskip0.2cm+{{\delta {\cal{H}}}\over{\delta\,{}^\#\pi ^{^{\underline{A}}}_a}}
\wedge{{\delta \omega}\over{\delta \underline{A}^a}}
-{{\delta \omega}\over{\delta\,{}^\#\pi ^{^{\underline{A}}}_a}}
\wedge {{\delta {\cal{H}}}\over{\delta \underline{A}^a}}\,.\cr }\eqno(\z)$$
As before, see (4.11,13), we have
$$\left\{\omega\,,{\cal{H}}\,\right\} 
=-\left\{{\cal{H}}\,,\omega\,\right\}\,,\eqno(\z)$$
and 
$$\left\{\sigma\wedge\omega\,,{\cal{H}}\,\right\} 
=\,\left\{\sigma\,,{\cal{H}}\,\right\}\wedge\omega  
+\sigma\wedge\left\{\omega\,,{\cal{H}}\,\right\} \,,\eqno(\z)$$
but in addition to (4.15), which remains valid, we can now
generalize it in order to take into account the contributions
due to the connections appearing in the covariant differentials.
Departing from the identity
$$u^0\,{\cal \L\/}_{e_{_0}} \underline{D}\,\omega ^a 
-\underline{D}\left(\,u^0\,{\cal \L\/}_{e_{_0}}
\omega ^a\,\right)\equiv\,u^0\,f^a{}_{bc}\,
F_{\bot}^b\wedge\omega ^c\,,\eqno(\z)$$
being $\omega ^a$ defined on the three--space, we get 
$$\left\{\,\underline{D}\,\omega ^a \,,{\cal{H}}\,\right\}
-\underline{D}\left\{ \omega ^a \,,{\cal{H}}\,\right\}
=\,f^a{}_{bc}\,\left\{\,\underline{A}^b 
\,,{\cal{H}}\,\right\}\wedge\omega ^c\,.\eqno(\z)$$
Now we require the stability of the constraint (5.7), i.e. we
impose that its time derivative vanishes. In terms of the
Poisson brackets we calculate its time evolution 
$$u^0\,{\it{l}}_{e_{_0}}\,{}^\#\pi ^{^{A_{\bot}}}_a 
=\,\left\{\,{}^\#\pi ^{^{A_{\bot}}}_a\,,{\cal{H}}\,\right\}
=\,u^0\,\underline{D}\,{}^\#\pi ^{^{\underline{A}}}_a \,,\eqno(\z)$$
and we put it equal to zero. The resulting stability condition 
$$\underline{D}\,{}^\#\pi ^{^{\underline{A}}}_a =\,0\eqno(\z)$$
is a secondary constraint. Further we require the time stability
of (5.17). Making use of (5.15), we find 
$$u^0\,{\it{l}}_{e_{_0}}\underline{D}\,{}^\#\pi ^{^{\underline{A}}}_a 
=\,\left\{\,\underline{D}\,{}^\#\pi ^{^{\underline{A}}}_a \,,
{\cal{H}}\,\right\} =-u^0\,f_{ab}{}^c\,A_{\bot}^b\,
\underline{D}\,{}^\#\pi ^{^{\underline{A}}}_c 
+2\,f_{(ab)}{}^c\,\underline{D}\left( u^0\,A_{\bot}^b
\,{}^\#\pi ^{^{\underline{A}}}_c\,\right)\,.\eqno(\z)$$
Assuming that we are dealing with a group whose structure
constants are antisymmetric in the first indices, eq.(5.18) 
reduces to 
$${\cal \L\/}_{e_{_0}}
\underline{D}\,{}^\#\pi ^{^{\underline{A}}}_a =\,0\,.\eqno(\z)$$
Thus, the constraint (5.17) is automatically stable and the search
for constraints is finished. The evolution of the system takes
place in a submanifold of the phase space defined by (5.7) and (5.17). 
The evolution equations are easily obtained as  
$$u^0\,{\it{l}}_{e_{_0}}\,\underline{A}^a 
=\,\left\{\,\underline{A}^a\,,{\cal{H}}\,\right\}
=\,u^0\,\pi ^{^{\underline{A}}}{}^{\,a} 
+\underline{D}\left(u^0\,A_{\bot}^a\,\right)\,,\eqno(\z)$$ 
$$u^0\,{\it{l}}_{e_{_0}} {}^\#\pi ^{^{\underline{A}}}_a 
=\,\left\{\,{}^\#\pi ^{^{\underline{A}}}_a \,,{\cal{H}}\,\right\}
=-\underline{D}\left(\,u^0\,{}^\#\underline{F}{}_a\,\right)
-u^0\,\overline{\eta}_{ab}{}^c\,A_{\bot}^b
\,{}^\#\pi ^{^{\underline{A}}}_c \,.\eqno(\z)$$
We rewrite (5.20,21) in the explicitly covariant form
$$\pi ^{^{\underline{A}}}{}^{\,a} =\,
{\it{l}}_{e_{_0}}\,\underline{A}^a 
-{1\over{u^0}}\,\underline{D}\left(u^0\,A_{\bot}^a\,\right)
=:F_{\bot}^a\,,\eqno(\z)$$ 
$${\cal \L\/}_{e_{_0}} {}^\#\pi ^{^{\underline{A}}}_a 
+{1\over{u^0}}\,\underline{D}\left(\,u^0\,
{}^\#\underline{F}{}_a\,\right)=\,0\,.\eqno(\z)$$
In addition, the remaining equation 
$$u^0\,{\it{l}}_{e_{_0}}\,A_{\bot}^a 
=\,\left\{\,A_{\bot}^a\,,{\cal{H}}\,\right\}
=\,\beta ^a \eqno(\z)$$
fixes the value of the Lagrange multiplier. 

Let us now compare our Hamiltonian equations with the usual 
four--dimensional Lagrangian equations $D\,{}^*F^a =\,0\,$. They
decompose into their normal and tangential parts as
$$D\,{}^*F^a =-\vartheta ^0\wedge\left[\,{\cal \L\/}_{e_{_0}}
{}^\#F_{\bot}^a +{1\over{u^0}}\,
\underline{D}\left(\,u^0\,{}^\#\underline{F}^a\,\right)\,\right]
-\underline{D}\,{}^\#F_{\bot}^a \,.\eqno(\z)$$
Since according to (5.22) the momentum $\pi ^{^{\underline{A}}}{}^{\,a}$
coincides with $F_{\bot}^a\,$, we see that the evolution
equation (5.23) corresponds to the normal part of (5.25), and the
constraint (5.17) to the tangential part. 
\bigskip\bigskip

\sectio{\bf Hamiltonian treatment of Gravitation}
\bigskip
At last we arrive at the main application of the general theory
established in sections 2--4. The nonlinear realization of the
Poincar\'e group and the Poincar\'e invariant foliation of spacetime
naturally related to it constitute the foundations of the
Hamiltonian approach to Gravity. In this context, we could of 
course choose as our starting point to derive the dynamical 
equations a very general action including quadratic terms. But
our purpose in this paper is to show how our formalism works,
and the best way to do so is to depart from the Einstein--Cartan
action, which gives rise to the standard Einstein equations of
GR. In fact, we will be able to derive them, see (7.8--14,18) 
below, thus showing that our approach comprises the usual 
Lagrangian one at the level of the field equations. However, 
our treatment is in some extent more restrictive, see section 8
below, and it has the additional virtue of distinguishing between 
those parts of the Einstein equations which are true evolution
equations and those which correspond to constraints on the phase 
space. Thus we depart from the Einstein--Cartan Lagrange density
with cosmological constant
$$L=-{1\over{2l^2}}\,R_\alpha{}^\beta\wedge\eta ^\alpha{}_\beta 
+{{\Lambda }\over{l^2}}\,\eta \,.\eqno(\z)$$
Let us reexpress the Einstein--Cartan term 
$$-{1\over{2l^2}}\,R_\alpha{}^\beta\wedge\eta ^\alpha{}_\beta 
=-{1\over{2l^2}}\,\left(\,2\,R^{0a}\wedge\eta _{0a} 
+R^{ab}\wedge\eta _{ab}\,\right)\,,\eqno(\z)$$
in terms of the threedimensional dynamical variables defined in
section 2. We point out once more that in our approach the
interactions are not mediated by the metric, which appears in
the theory as a trivial, nondynamical, anholonomic Kronecker
delta, but by the (nonlinear) connections. 

According to (2.12), the components of the fourdimensional
nonlinear Lorentz connections relate to the corresponding
$SO(3)$ quantities as 
$$\quad\Gamma ^{0a}=\,X^a\,,\quad
\Gamma ^{ab}=\,\epsilon ^{ab}{}_c\,A^c\,,\eqno(\z)$$
and accordingly, see (2.22,24,25),
$$R^{0a}=\,D\,X^a\quad\,,\qquad
R^{ab}=\,\epsilon ^{ab}{}_c\,{\cal{R}}^c\,.\eqno(\z)$$
On the other hand, from (E.1) follows 
$$\eta _{0a} =\,\overline{\eta}_a\quad\,,\qquad
\eta _{ab} =\,\vartheta ^0\wedge\overline{\eta}_{ab}\,.\eqno(\z)$$
Thus, substituting (6.4,5) into (6.2) and performing at the same
time the decomposition into the normal and tangential parts of 
the quantities involved as established in section 3, we get
$$-{1\over{2l^2}}\,R_\alpha{}^\beta\wedge\eta ^\alpha{}_\beta 
=-{1\over{l^2}}\,\vartheta ^0\wedge\left\{\,
\left[\,{\cal \L\/}_{e_{_0}}\underline{X}^a 
-{1\over{u^0}}\,\underline{D}\left( u^0\,X^a_{\bot}\,\right)\right]
\wedge\overline{\eta}_a +{1\over2}\,\epsilon ^{ab}{}_c
\,\underline{\cal{R}}^c\wedge\overline{\eta}_{ab}\,\right\}\,,\eqno(\z)$$
where we have made use of the definitions of the covariant Lie
derivative and the covariant tangential differential
respectively, namely
$$\eqalign{{\cal \L\/}_{e_{_0}}\underline{X}^a\,\,
&:=\,e_0\rfloor D\,\underline{X}^a 
=\,e_0\rfloor\left(\,d\,\underline{X}^a 
+\epsilon ^a{}_{bc}\,A^b\wedge\underline{X}^c\,\right)\cr 
\underline{D}\left( u^0\,X^a_{\bot}\,\right)
&:=\,\underline{d}\left( u^0\,X^a_{\bot}\,\right)
+\epsilon ^a{}_{bc}\,\underline{A}^b\,u^0\,X^c_{\bot}\,.\cr }
\eqno(\z)$$
According to (2.25,27), the tangential part of the curvature
present in (6.6) reads
$$\underline{\cal R}^a :=\,\underline{F}^a
-{1\over2}\,\epsilon ^a{}_{bc}\,\underline{X}^b
\wedge\underline{X}^c\quad\,,\qquad 
\underline{F}^a :=\,\underline{d}\,\underline{A}^a 
+{1\over2}\,\epsilon ^a{}_{bc}\,\underline{A}^b\wedge
\underline{A}^c \,.\eqno(\z)$$
For later convenience we also give here the structure of the
normal part of the curvature
$${\cal{R}}_{\bot}^a :=\,F_{\bot}^a 
-\epsilon ^a{}_{bc}\,X_{\bot}^b\,\underline{X}^c\quad\,,\qquad 
F_{\bot}^a :=\,{\it{l}}_{e_{_0}}\,\underline{A}^a
-{1\over{u^0}}\,\underline{D}\left(\,u^0\, A_{\bot}^a\,\right)
\,.\eqno(\z)$$
Despite it does not appear in the action, it will be present in
the field equations.

Due to the fact that the holonomic $SO(3)$ metric is the
Kronecker delta, we have ${\overline{\eta }}_{abc} 
={}^{\#}\left(\vartheta _a\wedge\vartheta _b\wedge\vartheta _c\,
\right) =\,\sqrt{\det (\delta _{mn})}\,\epsilon _{abc}=\,
\epsilon _{abc}\,$, see (E.1). Thus we identify this object with
the group constants of $SO(3)$ in order to simplify the
expression of the last term in (6.6). After integrating by parts
to make explicit the dependence on
${\cal \L\/}_{e_{_0}}\vartheta ^a\,$, we get the normal part of
the Lagrangian
$$L_{\bot}=-{1\over{l^2}}\,\left\{\,{1\over2}\,
\left[\,{\cal\L\/}_{e_{_0}}\underline{X}^a\wedge\overline{\eta}_a 
-\underline{X}^a \wedge {\cal \L\/}_{e_{_0}}\overline{\eta}_a\,
\right]-{1\over{u^0}}\,\underline{D}\left( u^0\,X^a_{\bot}\,\right)
\wedge\overline{\eta}_a +\vartheta ^a\wedge\,\underline{\cal{R}}_a 
-\Lambda\,\overline{\eta}\,\right\}\,.\eqno(\z)$$
Now we follow the steps of section 4, taking into account the
results of section 5 concerning the treatment of constrained
systems. First we define the momenta
$$\eqalign{^\#\pi ^{u^{^0}}
:=&\,{{\partial L_{\bot}}\over{\partial\left( 
{\it{l}}_{e_{_0}} u^0\,\right)}}=\,0\cr 
^\#\pi ^{\vartheta}_a 
:=&\,{{\partial L_{\bot}}\over{\partial\left( 
{\it{l}}_{e_{_0}}\vartheta ^a\,\right)}}
=-{1\over{2l^2}}\,\overline{\eta}_{ab}\wedge\underline{X}^b\cr 
^\#\pi ^{^{A_{\bot}}}_a 
:=&\,{{\partial L_{\bot}}\over{\partial\left( 
{\it{l}}_{e_{_0}} A_{\bot}^a\,\right)}}=\,0\cr 
^\#\pi ^{^{\underline{A}}}_a 
:=&\,{{\partial L_{\bot}}\over{\partial\left( 
{\it{l}}_{e_{_0}}\underline{A}^a\,\right)}}=\,0\cr 
^\#\pi ^{^{X_{\bot}}}_a 
:=&\,{{\partial L_{\bot}}\over{\partial\left( 
{\it{l}}_{e_{_0}} X_{\bot}^a\,\right)}}=\,0\cr 
^\#\pi ^{^{\underline{X}}}_a 
:=&\,{{\partial L_{\bot}}\over{\partial\left( 
{\it{l}}_{e_{_0}}\underline{X}^a\,\right)}}
=-{1\over{2l^2}}\,\overline{\eta}_a\,.\cr }\eqno(\z)$$
Observe that all of them are constraints. These are the primary
constraints which have to be included in the time evolution
operator by means of Lagrange multipliers, as in (5.10). It is 
interesting to notice that, acording to the last equation in
(6.11), the momenta conjugated to $\underline{X}^a$ are
proportional to the triads, i.e. $\pi ^{^{\underline{X}}}_a 
=-{1\over{2l^2}}\,\vartheta _a\,$. The canonical Hamiltonian, 
see (5.9), reads
$$\eqalign{{\cal{H}}_0 :=\,u^0\Bigl( &\hskip0.4cm
{\it{l}}_{e_{_0}} u^0\,\,{}^\#\pi ^{u^{^0}}
+
{\it{l}}_{e_{_0}}\vartheta ^a\wedge\,{}^\#\pi ^{\vartheta}_a
+
{\it{l}}_{e_{_0}} A_{\bot}^a\,\,{}^\#\pi ^{^{A_{\bot}}}_a\cr 
&+
{\it{l}}_{e_{_0}}\underline{A}^a\wedge\,{}^\#\pi ^{^{\underline{A}}}_a 
+
{\it{l}}_{e_{_0}} X_{\bot}^a\,\,{}^\#\pi ^{^{X_{\bot}}}_a 
+
{\it{l}}_{e_{_0}}\underline{X}^a\wedge\,{}^\#\pi ^{^{\underline{X}}}_a 
-L_{\bot}\Bigr)\,.\cr }\eqno(\z)$$ 
In order to construct the total Hamiltonian playing the role of
the time evolution operator, we apply the recipe established in
section 5. We put (6.12) into a form analogous to
(5.10) in the following sense. By adding and substracting
suitable terms, we rewrite (6.12), whenever possible, in terms
of covariant expressions. In this way, the remaining 
noncovariant contributions are rearranged into the terms
multiplied by $A_{\bot}^a$ and ${\it{l}}_{e_{_0}}A_{\bot}^a$
respectively. Then we substitute the factors multiplying the
primary constraints (6.11) by Lagrange multipliers 
$\beta ^i\,$, as in (5.10). (All the substituted terms depend on
time Lie derivatives, see the comment immediately after (5.10).)
The resulting total Hamiltonian 3--form reads
$$\eqalign{{\cal{H}}=&\,u^0\,\Bigl\{\,{1\over{l^2}}\left(\,
X^a_{\bot}\,\underline{D}\,\overline{\eta}_a 
+\vartheta _a \wedge\underline{\cal{R}}^a -\Lambda\,
\overline{\eta}\,\,\right)\cr 
&\hskip0.6cm -A_{\bot}^a\left[\,\underline{D}\,{}^\#\pi
^{^{\underline{A}}}_a +\overline{\eta}_{ab}{}^c\,
\left( X_{\bot}^b\,{}^\#\pi ^{^{X_{\bot}}}_c + 
\underline{X}^b\wedge {}^\#\pi ^{^{\underline{X}}}_c 
+\vartheta ^b\wedge {}^\#\pi ^{\vartheta}_c\,\right)
\,\right]\,\Bigr\}\cr 
&\hskip1.0cm +\beta ^0\,^\#\pi ^{u^{^0}}
+\beta _1 ^a\wedge\Bigl(\,^\#\pi ^{\vartheta}_a 
+{1\over{2l^2}}\,\overline{\eta}_{ab}\wedge\underline{X}^b\,\Bigr)
+\beta _2 ^a\,^\#\pi ^{^{A_{\bot}}}_a \cr 
&\hskip1.0cm+\beta _3 ^a\wedge {}^\#\pi ^{^{\underline{A}}}_a 
+\beta _4 ^a\,^\#\pi ^{^{X_{\bot}}}_a 
+\beta _5 ^a\wedge\Bigl(\,^\#\pi ^{^{\underline{X}}}_a 
+{1\over{2l^2}}\,\overline{\eta}_a\,\Bigr)\,.\cr }\eqno(\z)$$
The time evolution of any dynamical variable is calculable with
the help of the Poisson brackets, which take the general form
$$\eqalign{u^0\,{\it{l}}_{e_{_0}}\omega 
=\,\left\{\omega\,,{\cal{H}}\,\right\} &:=
{{\delta {\cal{H}}}\over{\delta\,{}^\#\pi ^{u^{^0}}}}
\,{{\delta \omega}\over{\delta u^0}}
\,\,\,\,\,-\,\,\,{{\delta \omega}\over{\delta\,{}^\#\pi ^{u^{^0}}}}
\,{{\delta {\cal{H}}}\over{\delta u^0}}\cr 
&\hskip0.2cm+{{\delta {\cal{H}}}\over{\delta\,{}^\#\pi ^{\vartheta}_a}}
\wedge{{\delta \omega}\over{\delta \vartheta ^a}}
-{{\delta \omega}\over{\delta\,{}^\#\pi ^{\vartheta}_a}}
\wedge {{\delta {\cal{H}}}\over{\delta \vartheta ^a}}\cr
&\hskip0.2cm+{{\delta {\cal{H}}}\over{\delta\,{}^\#\pi ^{^{A_{\bot}}}_a}}
\,{{\delta \omega}\over{\delta A_{\bot}^a}}
\,\,-\,{{\delta \omega}\over{\delta\,{}^\#\pi ^{^{A_{\bot}}}_a}}
\,{{\delta {\cal{H}}}\over{\delta A_{\bot}^a}}\cr
&\hskip0.2cm+{{\delta {\cal{H}}}\over{\delta\,{}^\#\pi ^{^{\underline{A}}}_a}}
\wedge{{\delta \omega}\over{\delta \underline{A}^a}}
-{{\delta \omega}\over{\delta\,{}^\#\pi ^{^{\underline{A}}}_a}}
\wedge {{\delta {\cal{H}}}\over{\delta \underline{A}^a}}\cr
&\hskip0.2cm+{{\delta {\cal{H}}}\over{\delta\,{}^\#\pi ^{^{X_{\bot}}}_a}}
\,{{\delta \omega}\over{\delta X_{\bot}^a}}
\,\,-\,{{\delta \omega}\over{\delta\,{}^\#\pi ^{^{X_{\bot}}}_a}}
\,{{\delta {\cal{H}}}\over{\delta X_{\bot}^a}}\cr
&\hskip0.2cm+{{\delta {\cal{H}}}\over{\delta\,{}^\#\pi ^{^{\underline{X}}}_a}}
\wedge{{\delta \omega}\over{\delta \underline{X}^a}}
-{{\delta \omega}\over{\delta\,{}^\#\pi ^{^{\underline{X}}}_a}}
\wedge {{\delta {\cal{H}}}\over{\delta \underline{X}^a}}\,.\cr }\eqno(\z)$$
They posses analogous properties to (5.11), namely
$$\left\{\omega\,,{\cal{H}}\,\right\} 
=-\left\{{\cal{H}}\,,\omega\,\right\}\,,\eqno(\z)$$
$$\left\{\sigma\wedge\omega\,,{\cal{H}}\,\right\} 
=\,\left\{\sigma\,,{\cal{H}}\,\right\}\wedge\omega  
+\sigma\wedge\left\{\omega\,,{\cal{H}}\,\right\} \,,\eqno(\z)$$
and
$$\left\{\,\underline{D}\,\omega ^a \,,{\cal{H}}\,\right\}
-\underline{D}\left\{ \omega ^a \,,{\cal{H}}\,\right\}
=\,\overline{\eta}^a{}_{bc}\,\left\{\,\underline{A}^b 
\,,{\cal{H}}\,\right\}\wedge\omega ^c\,,\eqno(\z)$$
compare with (5.12,13,15).

Making use of (6.14), we calculate the evolution equations of the
primary constraints (6.11). The stability requirement gives rise 
to four secondary constraints and to conditions on two Lagrange
multipliers respectively. In fact, we get the equations 
$$\eqalign{ u^0\,{\it{l}}_{e_{_0}}\,{}^\#\pi ^{u^{^0}}
=&-{1\over{l^2}}\,\left(\,\varphi ^{^{(0)}} + 
X^a_{\bot}\,\varphi ^{^{(3)}}_a\,\right) 
+A_{\bot}^a\,\varphi ^{^{(1)}}_a \cr 
u^0\,{\it{l}}_{e_{_0}}\,{}^\#\pi ^{^{A_{\bot}}}_a =&\hskip0.4cm 
u^0\,\varphi ^{^{(1)}}_a\cr 
u^0\,{\cal \L\/}_{e_{_0}}\,{}^\#\pi ^{^{\underline{A}}}_a 
=&-{1\over{l^2}}\,\varphi ^{^{(2)}}_a\cr
u^0\,{\cal \L\/}_{e_{_0}}\,{}^\#\pi ^{^{X_{\bot}}}_a 
=&-{{u^0}\over{l^2}}\,\varphi ^{^{(3)}}_a\cr
u^0\,{\cal \L\/}_{e_{_0}}\,\Bigl(\,^\#\pi ^{^{\underline{X}}}_a 
+{1\over{2l^2}}\,\overline{\eta}_a\,\Bigr)
=&-{1\over{l^2}}\,\overline{\eta}_{ab}\wedge\left(\,\beta _1^b
+u^0\,\underline{X}^b\,\right)\cr 
u^0\,{\cal \L\/}_{e_{_0}}\,\Bigl(\,^\#\pi ^{\vartheta}_a 
+{1\over{2l^2}}\,\overline{\eta}_{ab}\wedge\underline{X}^b\,\Bigr)
=&-{1\over{l^2}}\,\Bigl\{\,\Bigl[\,\beta^b_5 
-\underline{D}\left( u^0\,X^b_{\bot}\,\right)\,\Bigr]
\wedge\overline{\eta}_{ab} +u^0\,\Bigl(\,
\underline{\cal{R}}_a -\Lambda\,\overline{\eta}_a
\,\Bigr)\,\Bigr\}\,,\cr }\eqno(\z)$$ 
whose r.h.s.'s must vanish. The new secondary constraints in
(6.18) are defined as 
$$\eqalign{ \varphi ^{^{(0)}}:=&\,\vartheta _a 
\wedge\underline{\cal{R}}^a -\Lambda\,\overline{\eta}\cr 
\varphi ^{^{(1)}}_a:=&\,\underline{D}\,{}^\#\pi
^{^{\underline{A}}}_a +\overline{\eta}_{ab}{}^c\,
\left( X_{\bot}^b\,{}^\#\pi ^{^{X_{\bot}}}_c + 
\underline{X}^b\wedge {}^\#\pi ^{^{\underline{X}}}_c 
+\vartheta ^b\wedge {}^\#\pi ^{\vartheta}_c\,\right)\cr 
\varphi ^{^{(2)}}_a :=&\,\underline{D} (\,u^0\,
\vartheta _a\,) +u^0\,X_{\bot}^b\vartheta _b\wedge 
\vartheta _a\cr 
\varphi ^{^{(3)}}_a:=&\,\underline{D}
\,\overline{\eta}_a \,.\cr }\eqno(\z)$$
In addition, from (6.18e,f) also follows that the Lagrange
multipliers $\beta^a_1$ and $\beta^a_5$ satisfy respectively 
the equations 
$$\beta^a_1 =-u^0\,\underline{X}^a \,,\eqno(\z)$$
and 
$${1\over{u^0}}\,\Bigl[\,\beta^b_5 
-\underline{D}\left( u^0\,X^b_{\bot}\,\right)\,\Bigr]
\wedge\overline{\eta}_{ab} 
+\underline{\cal{R}}_a -\Lambda\,\overline{\eta}_a 
=\,0\,.\eqno(\z)$$
From the latter we find 
$$\beta^a_5 =\,\underline{D}\left( u^0\,X^a_{\bot}\,\right)
+u^0\,\left[\,\overline{\eta}^{abc}\left( e_b\rfloor
\underline{\cal{R}}_c\,\right) +{1\over2}\,\vartheta ^a
\left( e_b\rfloor {}^\#\underline{\cal{R}}^b\,\,\right) 
+{{\Lambda}\over2}\,\vartheta ^a\,\right]\,.\eqno(\z)$$
Let us now examine the consequences of the secondary constraints
(6.19). In view of the primary constraints (6.11), $\varphi ^{^{(1)}}_a$
reduces weakly to $\varphi ^{^{(1)}}_a =-{1\over{l^2}}\,
\vartheta _a\wedge\vartheta _b\wedge\underline{X}^b\,$. On the
other hand, the constraints $\varphi ^{^{(2)}}_a$ and 
$\varphi ^{^{(3)}}_a$ are not independent from each other. In
fact, the vanishing of $\varphi ^{^{(3)}}_a$ implies 
$\underline{D}\,\overline{\eta}_a =\,\underline{D}\,
\vartheta ^b\wedge\overline{\eta}_{ab}=\,0\,$. Substituting 
$\underline{D}\,\vartheta ^b =-\left(\underline{d}\log u^0 
+X^a_{\bot}\,\vartheta _a\,\right)\wedge\vartheta ^b$ as deduced
from $\varphi ^{^{(2)}}_a\,$, we get $\,{}^\#\left( e_a\rfloor
\underline{d}\log u^0 +X^b_{\bot}\,\vartheta _b\,\right)
=\,0\,$, thus proving that $\underline{d}\, \log u^0 
+ X^a_{\bot}\,\vartheta _a$ is a constraint by itself. Thus, 
$\underline{D}\,\vartheta ^a$ also vanishes. The latter
condition suffices to deduce $\varphi ^{^{(3)}}_a\,$. We
conclude that the new conditions imposed by the secondary
constraints (6.19) on the phase space manifold are
$$\eqalign{\vartheta _a\wedge\underline{\cal{R}}^a 
-\Lambda\,\overline{\eta} &=\,0\cr 
\vartheta _a\wedge\underline{X}^a &=\,0\cr 
\underline{d}\,\log u^0 +X_{\bot}^a\,\vartheta _a &=\,0\cr 
\underline{D}\,\vartheta ^a &=\,0\,.\cr }\eqno(\z)$$
Let us now require the stability of the secondary constraints (6.19).
Their time evolution calculated from the Hamiltonian (6.13) depends
on contributions which are weakly equal to zero in view of (6.11)
and (6.19) themselves --or equivalently (6.23)--, plus
additional terms which have to vanish in order to guarantee the 
stability. Firstly we verify that the stability of
$\varphi ^{(1)}_a$ is automatically fulfilled since 
$$u^0\,{\cal \L\/}_{e_{_0}}\,\varphi ^{(1)}_a\approx \,0\,.\eqno(\z)$$
(The $\approx $ term indicates that the equation holds weakly.) 
New conditions on the Lagrange multipliers are obtained when 
considering the stability of $\varphi ^{(2)}_a$ and 
$\varphi ^{(3)}_a\,$ in (6.19). We find 
$$u^0\,{\cal \L\/}_{e_{_0}}\,\varphi ^{(2)}_a\approx 
-u^0\,\left(\,\overline{\eta}_{ab}\wedge\beta _3^b 
-\underline{D}\,\beta ^1_a\,\right) 
-u^0\,\vartheta _a\wedge\left[\,\underline{d}\,
\left({{\beta ^0}\over{u^0}}\right) +\beta _4^b\,\vartheta _b 
+X_{\bot}^b\,\beta ^1_b\,\,\right]\,,\eqno(\z)$$
and
$$u^0\,{\cal \L\/}_{e_{_0}}\,\varphi ^{(3)}_a 
=-\left[\,\beta _3^b\wedge\vartheta _b\wedge\vartheta _a 
+\underline{D}\left(\,\overline{\eta}_{ab}\wedge\beta _1^b
\,\right)\,\right]\,.\eqno(\z)$$
Making use of the value of $\beta _1^a$ as given by (6.20), it
follows from (6.25,26) on the one hand
$$\underline{d}\,\left({{\beta ^0}\over{u^0}}\right) 
+\beta _4^a\,\vartheta _a -u^0\,X_{\bot}^a\,\underline{X}_a 
=\,0\,,\eqno(\z)$$
which guarantees that ${\cal \L\/}_{e_{_0}}\left(\,\underline{d}
\log u^0 +X_{\bot}^a\,\vartheta _a\,\right)=\,0\,$, in view 
of the meaning of $\beta ^0$ and $\beta _4^a$ deduced in (6.32)
below. On the other hand
$$\overline{\eta}_{ab}\wedge \beta _3^b 
+\underline{D}\left( u^0\,\underline{X}_a\,\right) =0\,,\eqno(\z)$$
from which we get 
$$\beta^a_3 =-\overline{\eta}^{abc}\left[ e_b\rfloor
\underline{D}\left(u^0\,\underline{X}_c\,\right)\,\right] 
-{1\over2}\,\vartheta ^a\left\{ e_b\rfloor {}^\#\left[\underline{D}
\left(u^0\,\underline{X}^b\,\right)\,\right]\,\right\}\,.\eqno(\z)$$
Finally we impose the stability condition on $\varphi ^{(0)}\,$,
see (6.19a). Taking into account the equations (6.20,21,28)
satisfied by $\beta _1^a\,$, $\beta _3^a $ and $\beta _5^a\,$,
we calculate 
$$u^0\,{\it{l}}_{e_{_0}}\,\varphi ^{(0)}\approx 
-\underline{d}\,u^0\wedge\varphi ^{(4)}
-\underline{d}\,\left( u^0\varphi ^{(4)}\,\right)\,,\eqno(\z)$$
with the new constraint $\varphi ^{(4)}$ defined as 
$$\varphi ^{(4)} :=\,\vartheta _a\wedge
{}^\#\underline{D}\,\underline{X}^a \,.\eqno(\z)$$
This constraint is stable. Thus, our search for the constraints
of the theory is finished. 

We end this section giving the evolution equations of the
canonical coordinates of the theory. They read
$$\eqalign{ u^0\,{\it{l}}_{e_{_0}} u^0\,\, =&\,\beta ^0\cr
u^0\,{\cal \L\/}_{e_{_0}}\vartheta ^a =&\,\beta _1^a\cr
u^0\,{\it{l}}_{e_{_0}} A_{\bot}^a =&\,\beta _2^a\cr
u^0\,F_{\bot}^a \hskip0.2cm=&\,\beta _3^a\cr 
u^0\,{\cal \L\/}_{e_{_0}} X_{\bot}^a =&\,\beta _4^a\cr
u^0\,{\cal \L\/}_{e_{_0}} \underline{X}^a =&\,\beta _5^a
\,.\cr }\eqno(\z)$$
(In (6.32d) we made use of definition (6.9).) The meaning of 
(6.32) will become clear below in view of the conditions 
established previously for the Lagrange multipliers. We do not
write down the evolution equations of the canonical momenta
(6.11) since most of them are zero constraints whose 
evolution equations have already been studied above, and those
of the nonvanishing momenta $^\#\pi ^{\vartheta}_a$ and 
$^\#\pi ^{^{\underline{X}}}_a$ are redundant with (6.32) due to 
the constraints (6.11) themselves. 
\bigskip\bigskip 

\sectio{\bf Comparison to the standard Einstein theory}
\bigskip 
At the end of section 5, we compared the Hamiltonian equations
of a Yang--Mills theory with the usual fourdimensional
Lagrangian equations derived from the same action and we saw
that they coincide. Here we will do the same with the gravitational
equations. The fourdimensional (Lagrangian) version of them,
obtained varying with respect to $\Gamma _\alpha{}^\beta$ and 
$\vartheta ^\alpha $ respectively, reads 
$$D\,\eta _{\alpha\beta}=\,0\,,\eqno(\z)$$
and
$${1\over2}\,\eta _{\alpha\beta\gamma}\wedge R^{\beta\gamma}
-\Lambda\,\eta _\alpha =\,0\,.\eqno(\z)$$
The curvature $R_\alpha{}^\beta$ in (7.2) as much as the covariant
differential in (7.1) are defined in terms of the fourdimensional
Lorentz connection $\Gamma _\alpha{}^\beta\,$. Eq.(7.1) establishes
the vanishing of the torsion. Thus we will substitute it by 
$$T^\alpha =\,0\,.\eqno(\z)$$
As a consequence of (7.3), the Lorentz connection reduces to the
Christoffel symbol$^{(6,7)}$
$$\Gamma ^{\{\}}_{\alpha\beta}:=\,
e_{[\alpha }\rfloor d\,\vartheta _{\beta ]} -{1\over2}
\left( e_\alpha\rfloor e_\beta\rfloor d\,\vartheta ^\gamma\right) 
\vartheta _\gamma\,,\eqno(\z)$$ 
and (7.2) coincides with the standard Einstein vacuum equations
with cosmological constant defined on a Riemannian space.

Let us now decompose the time and space components of the
Lagrangian equations into their normal and tangential parts
respectively, according to the foliation procedure of section 3.
For the torsion equation (7.3) we find, see (2.23) 
$$\eqalign{ 0=&\,T^0:=\,d\,\vartheta ^0 +\vartheta _a\wedge X^a
=-\vartheta ^0\wedge\left(\underline{d}\,\log u^0 
+X^a_{\bot}\vartheta _a\,\right)+\vartheta _a\wedge
\underline{X}^a \cr 
0=&\,T^a :=\,D\,\vartheta ^a +\vartheta ^0\wedge X^a 
=\,\vartheta ^0\wedge\left( {\cal \L\/}_{e_{_0}}\vartheta ^a 
+\underline{X}^a\,\right) +\underline{D}\,\vartheta ^a\,.\cr }
\eqno(\z)$$
On the other hand, the Einstein equations (7.2) decompose as 
$$0=\,{1\over2}\,\eta _{0\beta\gamma}\wedge R^{\beta\gamma}
-\Lambda\,\eta _0 =-\vartheta ^0\wedge\left(\vartheta _a\wedge
{\cal{R}}_{\bot}^a\,\right) +\left(\,\vartheta _a\wedge
\underline{\cal{R}}^a -\Lambda\,\overline{\eta}\,\,\right)\,,\eqno(\z)$$
and 
$$\eqalign{0&=\,{1\over2}\,\eta _{a\beta\gamma}\wedge R^{\beta\gamma}
-\Lambda\,\eta _a \cr 
&=-\vartheta ^0\wedge \Biggl\{\Bigl[\,
{\cal \L\/}_{e_{_0}}\,\underline{X}^b 
-{1\over{u^0}}\,\underline{D}\left( u^0\,X^b_{\bot}\,\right)
\,\Bigr]\wedge\overline{\eta}_{ab} +\underline{\cal{R}}_a 
-\Lambda\,\overline{\eta}_a\,\Biggr\} -\overline{\eta}_{ab}
\wedge\underline{D}\,\underline{X}^b\,.\cr }\eqno(\z)$$ 
Our task now is to compare the set of Lagrangian equations (7.5--7)
with our Hamiltonian ones. In order to do it, we will rearrange
the results of section 6, in particular equations (6.20,21,28) 
{\it cum} (6.32), into more explicit expressions, see below. As
a general result, the tangential parts of all the Lagrangian
equations vanish due to the secondary constraints we have found.
In fact, the dynamical meaning of the constraints becomes 
transparent when comparing with (7.5--7). We read out from
(6.23) the following conditions. For $T^0\,$, compare (6.23b)
and (7.5a):
$$\vartheta _a\wedge\underline{X}^a =\,0\,,\eqno(\z)$$
for $T^a\,$, compare (6.23d) and (7.5b):
$$\underline{D}\,\vartheta ^a =\,0\,,\eqno(\z)$$
and for the time component of the Einstein equations, compare
(6.23a) and (7.6): 
$$\vartheta _a\wedge\underline{\cal{R}}^a 
-\Lambda\,\overline{\eta}=\,0\,.\eqno(\z)$$
Furthermore, from the constraint (6.31) follows immediately the 
wanishing of the tangential part of the space components of
the Einstein equations (7.7), namely
$$\overline{\eta}_{ab}\wedge\underline{D}\,\underline{X}^b 
=\,0\,.\eqno(\z)$$ 
In short, the constraints (7.8--11) are the tangential parts of
the Lagrange equations (7.5--7). Sequently let us pay attention
to the normal parts. With the only exception of that of $T^0\,$,
see (7.5a), which appears as the constraint (6.23c):
$$\underline{d}\,\log u^0 +X^a_{\bot}\vartheta _a =\,0\,,
\eqno(\z)$$
the normal parts are obtained from the evolution equations (6.32)
and the conditions (6.20,21,28) on the Lagrange multipliers. Putting
together (6.32b) and (6.20), it follows
$${\cal \L\/}_{e_{_0}}\vartheta ^a +\underline{X}^a =\,0\,,\eqno(\z)$$
compare with the normal part of (7.5b). On the other hand,
substituting (6.32f) into (6.21), we get 
$$\Bigl[\,{\cal \L\/}_{e_{_0}}\,\underline{X}^b 
-{1\over{u^0}}\,\underline{D}\left( u^0\,X^b_{\bot}\,\right)
\,\Bigr]\wedge\overline{\eta}_{ab} +\underline{\cal{R}}_a 
-\Lambda\,\overline{\eta}_a =\,0\,,\eqno(\z)$$
which corresponds to the normal part of (7.7). Finally, from
(6.28) and (6.32d) we find
$$u^0\,\overline{\eta}_{ab}\wedge F_{\bot}^b 
+\underline{D}\left( u^0\,\underline{X}_a\,\right) 
=\,0\,,\eqno(\z)$$
which, after an obvious manipulation, gives rise to
$$\vartheta _a\wedge\vartheta _b\wedge F_{\bot}^b 
+\underline{d}\,\log u^0\wedge\overline{\eta}_{ab}
\wedge\underline{X}^b -\overline{\eta}_{ab}\wedge
\underline{D}\,\underline{X}^b =\,0\,.\eqno(\z)$$
Taking into account (7.11,12) and the fact that (7.8) implies 
$e_{[\,a}\rfloor\underline{X}_{b\,]}=\,0\,$, the contraction of
(7.16) with $e_a$ leads to 
$$\vartheta _a\wedge\left( F_{\bot}^a -\overline{\eta}^a{}_{bc}
\,X_{\bot}^b\,\underline{X}^c\,\right) =\,0\,.\eqno(\z)$$ 
According to definition (6.9), the previous equation may be
rewritten as
$$\vartheta _a\wedge {\cal{R}}_{\bot}^a =\,0\,,\eqno(\z)$$
which reproduces the normal part of (7.6). Thus, we were able to
reproduce all the Einstein equations from our Hamiltonian
approach. 
\bigskip\bigskip

\sectio{\bf The SO(3) formulation of Einstein's equations}
\bigskip 
Once we have proven that the usual Lagrangian equations are a
consequence of our treatment, let us write down the Hamiltonian 
evolution equations in a more suitable and simple form, which
shows more clearly their physical meaning. As a consequence, we 
will see that our equations (8.7--10) below are more restrictive
than those of the Einstein theory. In the first place, we 
reexpress (7.12) as 
$$X_{\bot}^a =-\left( e^a\rfloor\underline{d}\,\log u^0\,\right)
\,.\eqno(\z)$$ 
From (7.8) together with (7.13) it follows that 
$$A_{\bot}^a =-{1\over2}\,\overline{\eta}^{abc}
\left( e_b\rfloor {\it{l}}_{e_{_0}}\vartheta _c\,\right)\,,\eqno(\z)$$
and
$$\underline{X}^a =-{\cal \L\/}_{e_{_0}}\vartheta ^a 
=-{1\over2}\left[\,{\it{l}}_{e_{_0}}
\vartheta ^a +\vartheta _b \left( e^a\rfloor {\it{l}}_{e_{_0}}
\vartheta ^b\,\right)\,\right]\,,\eqno(\z)$$
and on the other hand we deduce from (7.9)  
$$\underline{A}^a =-{1\over2}\,\overline{\eta}^{abc}
\left[ e_b\rfloor\underline{d}\,\vartheta _c -{1\over2}
\left( e_b\rfloor e_c\rfloor \underline{d}\,\vartheta ^d\right) 
\vartheta _d\,\right]\,.\eqno(\z)$$ 
Equations (8.1--4) reproduce (7.4) decomposed into its
constitutive parts, i.e. they ensure that the torsion vanish and
that the nonlinear connection reduces to the Christoffel symbol. The 
equations (8.1--4) are equivalent to those (7.8,9,12,13) from
which we derived them. In their original form, we see that three
of them, namely (7.8,9,12) are constraints, whereas the fourth
one (7.13) is an evolution equation. In the usual interpretation
of GR, all these conditions are accepted {\it a priori} to hold,
as a constitutive part of the Riemannian geometrical background.
In this paper we will not enter into the discussion of this
point and its consequences. We will do it in a work in course.
But we point out that the ignorance of the dynamical character
of the vanishing of the torsion is a source of open problems,
mostly when one attempts to quantize Gravity.

Let us now look at the remaining dynamical equations. Taking 
the covariant differential of (7.9), it follows 
$\overline{\eta}_{ab}\wedge\underline{F}^b =\,0\,$, which
gives rise to $\vartheta _a\wedge\,^\#\underline{F}^a =\,0\,$. 
On the other hand, as pointed out above, (7.8) implies 
$e_{[\,a}\rfloor\underline{X}_{b\,]}=\,0\,$. From both results, 
together with definition (6.8), we conclude that $\vartheta _a\wedge
\,^\#\underline{\cal{R}}^a =\,0\,$. Similarly, from (7.10)
we obtain $\left( e_a\rfloor\,^\#\underline{\cal{R}}^a\,\right) 
=\,\Lambda\,$. Making use of these results, (6.22) reduces to 
$$\beta _5^a =\,\underline{D}\left( u^0\,X_{\bot}^a\right) 
+u^0\,{}^\#\underline{\cal{R}}^a\,.\eqno(\z)$$
It is also easy to prove that (6.29) reduces to
$$\beta _3^a =\,u^0\,\overline{\eta}^a{}_{bc}
\,X_{\bot}^b\,\underline{X}^c -u^0\,^\#\left(\underline{D}
\,\underline{X}^a\right)\,.\eqno(\z)$$
Substituting these expressions into (6.32d,f) and taking the 
definition (6.9) into account, we obtain 
$${\cal \L\/}_{e_{_0}}\underline{X}^a
-{1\over{u^0}}\,\underline{D}\left(\,u^0\,X^a_{\bot}\right)
=\,^\#\underline{\cal{R}}^a\,,\eqno(\z)$$
and
$${\cal{R}}_{\bot}^a =-\,^\#\left(\underline{D}\,\underline{X}^a\right) 
\,.\eqno(\z)$$
In addition, we have eqs.(7.10) and (7.18), namely
$$\vartheta _a\wedge\underline{\cal{R}}^a 
=\,\Lambda\,\overline{\eta}\,,\eqno(\z)$$
and
$$\vartheta _a\wedge{\cal{R}}_{\bot}^a =\,0\,.\eqno(\z)$$
Eq.(8.10) is the trace of (8.8) with the constraint (6.31) taken
into account. This completes the dynamical information derived
from section 6. What is important to be noticed here is that,
whereas (8.10) is present in the Lagrangian equations, (8.8) is
not. It is a further restriction to be added to the standard
Einstein theory.

Eqs.(8.7,8) represent the time evolution of the tangential parts
$\underline{X}^a =-{\cal \L\/}_{e_{_0}}\vartheta ^a $ of the boost
vector and $\underline{A}^a $ of the $SO(3)$ connection, see
(6.9), respectively. Let us put them together into a
fourdimensional formula in order to show them in their simplest
form. According to (3.16b), we have
$$^*{\cal{R}}^a =\,\vartheta ^0\wedge {}^\#\underline{\cal{R}}^a 
-{}^\# {\cal{R}}_{\bot}^a \,,\eqno(\z)$$
and on the other hand, see (3.19),
$$D\,X^a =\,\vartheta ^0\wedge\left[ {\cal \L\/}_{e_{_0}}\underline{X}^a
-{1\over{u^0}}\,\underline{D}\left(\,u^0\,X^a_{\bot}\right)\,\right]
+\underline{D}\,\underline{X}^a \,.\eqno(\z)$$
In view of (8.11,12), eq.(8.7) and the dual of (8.8) rearrange
into 
$$D\,X^a -{}^*{\cal{R}}^a =\,0\,.\eqno(\z)$$
Analogously, (8.9) together with (8.10) take the fourdimensional form 
$$\vartheta _a\wedge {\cal{R}}^a -\Lambda\,\eta _{_0} =\,0\,,\eqno(\z)$$
being $\eta _{_0}:= e_{_0}\rfloor\eta\,$, see (E.1). Since the
torsion vanishes according to (8.1--4), eqs.(8.13,14)
are the condensed form of the Hamiltonian Einstein equations on
a Riemannian spacetime.
\bigskip\bigskip

\sectio{\bf Relationship with Ashtekar variables}\bigskip
Finally, let us briefly discuss how our variables relate to
those of Ashtekar$^{(9,16)}$. The Ashtekar description derives from 
the usual $SO(3)$ ADM scheme with dynamical variables 
$\left( E_a{}^i\,,K_i{}^a\right)$ through a quasi Legendre 
transformation of the form
$$\eqalign{ E_a{}^i&\longrightarrow E_a{}^i\cr 
K_i{}^a&\longrightarrow A_i{}^a =\,\Gamma _i{}^a 
+\beta\,K_i{}^a\,,\cr }\eqno(\z)$$ 
where $\Gamma _i{}^a$ is an $SO(3)$ connection compatible with
$E_a{}^i\,$, and $\beta$ is a constant to be fixed
later. In this way, the Gauss constraint becomes 
$$\nabla _i\,E_a{}^i =\,0\,,\eqno(\z)$$
being the covariant derivative $\nabla _i$ constructed in terms
of the $SO(3)$ connection $A_i{}^a\,$. The vectorial constraint
takes the form
$$F_{ij}{}^a\,E_a{}^j =\,0\,,\eqno(\z)$$
with $F_{ij}{}^a$ as the usual $SO(3)$ field strength tensor. 
Finally, the scalar constraint becomes 
$$-\zeta\,\epsilon _{abc}\, E^a{}_i E^b{}_j F^{ijc}
+2\,\left(\zeta -{1\over{\beta ^2}}\right) 
E_{[\,a}{}^i E_{\,b\,]}{}^j\,\left( A_i{}^a 
-\Gamma _i{}^a\,\right)\left( A_j{}^b 
-\Gamma _j{}^b\,\right) =\,0\,,\eqno(\z)$$
where $\zeta$ stands for the signature, corresponding in
particular $\zeta=-1$ to the Lorentzian one considered by us. 

Let us now compare (9.2--4) with our results. In order to do
so, we will express the relevant equations obtained in section 6 in 
terms of components, making use of the notation 
$$\vartheta ^a =\,e^a{}_i\,d\, x^i\quad\,,\qquad 
\underline{X}^a =\,\underline{X}_i{}^a\,d\, x^i\quad\,,\qquad 
\underline{A}^a =\,\underline{A}_i{}^a\,d\, x^i\,.\eqno(\z)$$
The constraint (6.19d) gives rise to 
$$^\#\Bigl(\underline{D}\,\overline{\eta}_a\,\Bigr)
=\,\underline{D}{}_i e_a{}^i =\,0\,,\eqno(\z)$$
which coincides with the Gauss constraint (9.2) if we identify 
$e^a{}_i$ with $E^a{}_i\,$. On the other hand, taking
the covariant differential of (6.23d) we get 
$$\underline{D}\wedge\underline{D}\,\vartheta ^a 
\equiv\,\overline{\eta}^a{}_b\wedge\underline{F}^b=\,0\,.\eqno(\z)$$
From (9.7) follows the vectorial costraint
$$e^a{}_i\,{}^\#\Bigl(\,\overline{\eta}_{ab}\wedge\underline{F}^b\,\Bigr) 
=\,\underline{F}_{ij}{}^a\,e_a{}^j =\,0\,,\eqno(\z)$$
compare with (9.3). Finally, we develop our scalar constraint 
(6.23a) as 
$$^\#\Bigl(\vartheta _a\wedge\underline{\cal{R}}^a
-\Lambda\,\overline{\eta}\,\Bigr) =\,{1\over 2}\,
\left(\epsilon _{abc}\, e^a{}_i e^b{}_j\underline{F}^{ijc}
-2\,e_{[\,a}{}^i e_{\,b\,]}{}^j\,\underline{X}_i{}^a\,
\underline{X}_j{}^b\,\right)-\Lambda =\,0\,.\eqno(\z)$$
Taking $\underline{X}_i{}^a$ proportional to $\left( A_i{}^a 
-\Gamma _i{}^a\,\right)\,$, eq.(9.9) with $\Lambda =\,0$ coincides
with (9.4), with the Lorentzian signature $\zeta =-1\,$, in the
limit $\beta = 1$ suggested by Barbero$^{(16)}$. Furthermore, the
standard complex Ashtekar variables corresponding to the choice
$\beta = i$ also relate to ours in a simple way as follows. Let
us perform the transformation 
$$\underline{A}^a\longrightarrow \tilde{\underline{A}}^a 
=\,\underline{A}^a +i\,\underline{X}^a\,.\eqno(\z)$$
The field strength constructed with the complex $SO(3)$
connection $\tilde{\underline{A}}^a\,$, in terms of the
original real variables reads 
$$\tilde{\underline{F}}^a =\,\underline{\cal{R}}^a 
+i\,\underline{D}\,\underline{X}^a\,,\eqno(\z)$$
see (6.8). Taking now the constraints (7.8,9,11) into
account, we verify that 
$$\underline{D}\,\overline{\eta}_a
=\,\tilde{\underline{D}}\,\overline{\eta}_a
\,,\qquad
\overline{\eta}_{ab}\wedge\underline{F}^b
=\,\overline{\eta}_{ab}\wedge\tilde{\underline{F}}^b
\,,\qquad
\vartheta _a\wedge\underline{\cal{R}}^a
=\,\vartheta _a\wedge\tilde{\underline{F}}^a\,.\eqno(\z)$$
The expressions with tilde depend on the complex connection, see
(9.10). In terms of them, the constraints (9.6,8,9) take
the usual Ashtekar form, namely 
$$^\#\left(\underline{D}\,\overline{\eta}_a\,\right)
=\,\tilde{\underline{D}}{}_i e_a{}^i =\,0\,,\eqno(\z)$$
$$e^a{}_i\,{}^\#\Bigl(\,\overline{\eta}_{ab}\wedge\underline{F}^b\,\Bigr) 
=\,\tilde{\underline{F}}_{ij}{}^a\,e_a{}^j =\,0\,,\eqno(\z)$$
and 
$$^\#\Bigl(\vartheta _a\wedge\underline{\cal{R}}^a
-\Lambda\,\overline{\eta}\,\Bigr) =\,{1\over 2}\,
\epsilon _{abc}\, e^a{}_i e^b{}_j\tilde{\underline{F}}^{ijc} 
-\Lambda =\,0\,.\eqno(\z)$$
We point out that the physical meaning of the change (9.1) in the 
Ashtekar approach becomes evident in relation with the dynamics
of the $X^a$ vectors associated to the boosts. 
\bigskip\bigskip

\noindent{\bf Conclusions}\bigskip
Taking advantage of the fact that a particular nonlinear 
realization of the Poincar\'e group defines a natural time 
direction, we performed a Poincar\'e invariant spacetime
foliation, and we constructed a Hamiltonian formalism adapted 
to the local spacetime symmetry. We identified the gravitational
dynamical variables to be nonlinear connections (differential 1--forms) 
with $SO(3)$ indices corresponding to the classification subgroup. 

From the Hamiltonian evolution equations corresponding to the 
Einstein--Cartan action we reproduced the standard Lagrangian 
field equations of GR, but we also proved that the Hamiltonian
ones are more restrictive, and we obtained the complete set of
constraints. Ashtekar variables were identified with the 
{\it natural} dynamical coordinates of our nonlinear Hamiltonian
description of Gravity.

It would be interesting to study more general actions. In fact,
as a consequence of the nonlinear approach, a large number of 
Poincar\'e invariants exist with respect to the explicit $SO(3)$
classification subgroup, which are not expressible in the 
fourdimensional geometrical language, and accordingly we have at
our disposal additional invariant terms which make more flexible
the choice of gravitational actions.
\bigskip\bigskip

\centerline {\bf Acknowledgements}\bigskip
We cordially thank Prof. Manuel de Le\'on for his valuable
suggestions and help in several hard mathematical questions, 
Dr. Fernando Barbero for his contribution to our understanding 
of the link between the nonlinear gauge fields and Ashtekar 
variables, and Prof. Friedrich Wilhelm Hehl for advising us on
his transparent treatment of gravitational gauge theories, and 
for constant interest in our work.
\bigskip\bigskip\bigskip
\centerline {\bf APPENDICES}
\bigskip\bigskip
\noindent{\bf{A.--Coset realizations of symmetry groups}}
\bigskip 
In this appendix we briefly summarize the nonlinear coset
realization procedure$^{(4)}$ which constitutes the basis of the
particular application of section 2 and thus of the whole 
present work.

Let $G=\{g\}$ be a Lie group including a subgroup $H=\{h\}$ 
whose linear representations $\rho (h)$ are known, acting on 
functions $\psi $ belonging to a linear representation space of 
$H$. The elements of the quotient space $G/H$ are equivalence 
classes of the form $gH=\{gh_1\,,gh_2\,...\,gh_n\}\,$, 
and they constitute a complete partition of the group space. 
We call the elements of the quotient space cosets to the left 
(right) of $G$ with respect to $H$. Since we deal with Lie 
groups, the elements of $G/H$ are labeled by continuous
parameters, say $\xi$. We represent the elements of $G/H$ by
means of the coset indicators $c(\xi )\,$, parametrized by 
the coset parameters $\xi\,$, playing the role of 
a kind of coordinates. The nonlinear coset realizations 
are based on the action of the group on $G/H$, i.e., on a 
partition of its own space. An arbitrary element $g\epsilon G$ 
acts on $G/H$ transforming a coset into another, that is
$$\eqalign{g:\, G/H&\rightarrow G/H\cr
c\,(\xi )&\rightarrow c\,(\xi ')\,,\cr }\eqno(A.1)$$
according to the general law
$$g\,c\,(\xi\,)=\, c\,(\xi ')\, h\left(\xi\,, g\right)\,.\eqno(A.2)$$
The elements $h\left(\xi\,,g\right)$ which appear in (A.2) belong
to the subgroup $H$, that we will call in the following the
{\it{classification subgroup}}, since the elements $g$ of the
whole group $G$ considered in (A.2) act nonlinearly on the
representation space of the classification subgroup $H$ according to
$$\psi '=\,\rho\left(h\left(\xi\,, g\right)\right)\psi\,,\eqno(A.3)$$
where $\rho$, as mentioned above, is a linear representation
of $H$ in the space of the matter fields $\psi$. Therefore, the
action of the total group $G$ projects on the representations of
the subgroup $H$ through the dependence of $h\left(\xi\,,g\right)$ 
in (A.2) on the group element $g$, as given by eq.(A.3). The action of
the group is realized on the couples $\left(\xi\,,\psi\right)$.
It reduces to the usual linear action of $H$ when we take in
particular for $g$ in (A.2) an element of $H$.

In order to define a covariant differential transforming like 
(A.3) under local transformations, we need a suitable nonlinear 
connection. We define it as
$$\Gamma:=\, c^{-1}{\rm {\cal{D}}}c\,,\eqno(A.4)$$ 
where the covariant differential on the coset space is defined as
$${\rm {\cal{D}}}c:=\,\left(d\,+\Omega\,\right)c\,,\eqno(A.5)$$ 
with the ordinary linear connection $\Omega$ of the whole group G 
transforming as
$$\Omega '=\, g\,\Omega \,g^{-1}+g\,d\,g^{-1}\,.\eqno(A.6)$$
It is easy to prove that the nonlinear gauge field $\Gamma$
defined in (A.4) transforms as
$$\Gamma '=\,h \Gamma h^{-1} +h d\, h^{-1}\,,\eqno(A.7)$$ 
thus allowing to define the nonlinear covariant differential 
operator 
$${\bf D}:=\,d\, +\Gamma\,.\eqno(A.8)$$
One can read out from (A.7) that only the components of $\Gamma$
related to the generators of $H$ behave as true connections, 
transforming inhomogeneously, whereas the components of $\Gamma$
over the generators associated with the cosets $c$ transform as 
tensors with respect to the subgroup $H$ notwithstanding their 
nature of connections.
\bigskip\bigskip 

\noindent{\bf{B.--The Poincar\'e group in terms of boosts, rotations and
space and time translations}}\bigskip 
The calculations leading to the results of section 2 rest on the
decomposition of the generators of the Poincar\'e group presented
in this Appendix. In particular, we made use of the commutation
relations (B.6). 

In the fourdimensional notation, the Lorentz generators 
$L_{\alpha\beta}$ and the translational generators
$P_\alpha\hskip0.3cm (\alpha\,,\beta =\,0...3)$ of the Poincar\'e
group satisfy the commutation relations
$$\eqalign{\left[L_{\alpha\beta }\,,L_{\mu\nu }\right]\hskip0.10cm&=
-i\,\left( o_{\alpha [\mu } L_{\nu ]\beta}   
         - o_{\beta [\mu }L_{\nu ]\alpha }\right)\,,\cr 
\left[L_{\alpha\beta }\,, P_\mu\,\right]\hskip0.10cm&=
\,\,\,\,i\,o_{\mu [\alpha }P_{\beta ]}\,,\cr 
\left[ P_\alpha\,, P_\beta\,\right]\hskip0.10cm&
=\,\,\,0\,.\cr}\eqno(B.1)$$
We choose the invariant metric tensor to be 
$$o_{\alpha\beta}:=\,diag(-\,+\,+\,+\,)\,.\eqno(B.2)$$
We can decompose the generators in such a way that
$$\beta ^{\alpha\beta}L_{\alpha\beta}=\,2\beta ^{a0}L_{a0}
+\beta ^{ab}L_{ab}\,,\eqno(B.3)$$
with $a\,,b$ running from 1 to 3. Let us define 
$$S_a :=-\epsilon _a{}^{bc} L_{bc}\,,\eqno(B.4)$$
$$K_a :=\,2\,L_{a0}\,.\eqno(B.5)$$
The generators (B.4) are those of the $SO(3)$ group, and (B.5)
correspond to the boosts. In terms of them, and taking (B.2) into
account, the commutation relations (B.1) transform into 
$$\eqalign{\left[ S_a\,,S_b\,\right]&=-i\,\epsilon _{abc}S_c\,,\cr
\left[ K_a\,, K_b\,\right]&=\,\,\,\,i\,\epsilon _{abc}S_c\,,\cr
\left[ S_a\,,K_b\,\right]&=-i\,\epsilon _{abc}K_c\,,\cr
\left[ S_a\,,P_0\,\right]&=\,\,\,\,0\,,\cr
\left[ S_a\,,P_b\,\right]&=-i\,\epsilon _{abc}P_c\,,\cr
\left[ K_a\,,P_0\,\right]&=\,\,\,\,i\,P_a\,,\cr
\left[ K_a\,,P_b\,\right]&=\,\,\,\,i\,\delta _{ab}P_0\,,\cr
\left[ P_a\,,P_b\,\right]&=\,
\left[ P_a\,,P_0\,\right] =\,
\left[ P_0\,,P_0\,\right] =\,0\,.\cr}\eqno(B.6)$$
Thus we can rewrite (B.3) as
$$\beta ^{\alpha\beta}L_{\alpha\beta}=\,\xi ^a K_a +\theta ^a
S_a \,,\eqno(B.7)$$
with the new coefficients used in section 2 defined in an
obvious way.

The following well known formulae will be useful for the
audacious reader who wants to reproduce the calculations.
$$\eqalign{\epsilon _{abc}\epsilon ^{mns}&=\,2\,\left( 
\delta ^m_{[a}\delta ^n_{b]}\delta ^s_c +
\delta ^s_{[a}\delta ^m_{b]}\delta ^n_c +
\delta ^n_{[a}\delta ^s_{b]}\delta ^m_c\right)\,,\cr
\epsilon _{abc}\epsilon ^{mnc}&=\,2\,\delta ^m_{[a}\delta ^n_{b]}\,,\cr
\epsilon _{abc}\epsilon ^{mbc}&=\,2\,\delta ^m_a\,,\cr
\epsilon _{abc}\epsilon ^{abc}&=\,3!\,.\cr}\eqno(B.8)$$
$$\eqalign{e^{-A} B e^A&=\,B-[A\,,B\,]
+{1\over{2!}}\,[A\,,[A\,,B\,]\,]-...\cr 
e^{-\chi A}d\,e^{\chi A}&=\,d\,\chi A
-{1\over{2!}}\,[\chi A\,,d\,\chi A\,]
+{1\over{3!}}\,[\chi A\,,[\chi A\,,d\,\chi A\,]\,]-...\cr }
\eqno(B.9)$$
$$e^{i\,(\lambda ^a +\delta\lambda ^a ) K_a}
=\,e^{i\,\lambda ^a K_a}\left( 1 +e^{-i\,\lambda ^a K_a}
\delta e^{i\,\lambda ^a K_a}\,\right)\,.\eqno(B.10)$$
The choice of signs in (2.11,12) is conventional. We wanted to be
consistent with our previous work$^{(3)}$ and at the same time 
we attempted to reproduce the usual sign conventions in the
definition of $SO(3)$ covariant differentials and curvature.
\bigskip\bigskip 

\noindent{\bf C.--The Hamiltonian formalism}
\bigskip 
Here we present an alternative deduction of the Hamiltonian
formalism. We will not pay attention to the constraints since we
are only interested in showing that the Lie derivatives obtained
in the abbreviated way of section 4 coincide with those derived
in the standard approach sketched in this Appendix. Let us
consider a Lagrangian density 
$$L=\,L\left(\,u^0\,,d\,u^0\,,\alpha\,,d\,\alpha\,\right)
\,,\eqno(C.1)$$
being $\alpha$ a p--form. We include in (C.1) the explicit 
dependence on $u^0$ in order to take into account the 
applicability of the formalism to Gravity, as it is done in 
section 6. In the absence of a gravitational action, $u^0$
belongs to the geometrical background and does not play any
dynamical role. We decompose the variables $\alpha$ into their
normal and tangential parts, see (3.16,19), and accordingly we 
rewrite the Lagrangian density as (4.3), with 
$$L_{\bot}=\,L_{\bot}\left(\,u^0\,,\,\underline{d}\,u^0\,,\,
\alpha _{\bot}\,,\,\underline{\alpha}\,,\,\underline{d}
\,\alpha _{\bot}\,,\,\underline{d}\,\underline{\alpha}\,,\,
{\it{l}}_{e_{_0}}\underline{\alpha}\,\right)\,.\eqno(C.2)$$
The variation 
$$\eqalign{\delta L=&\,\vartheta ^0\wedge\Biggl\{ 
\delta u^0\,\Biggl[ {{\partial L_{\bot}}\over{\partial u^0}}
-{1\over{u^0}}\,\underline{d}\,\biggl( u^0\,{{\partial L_{\bot}}
\over{\partial\bigl(\underline{d}\,u^0\bigr)}}\biggr)
-{1\over{u^0}}\,\biggl( {\it{l}}_{e_{_0}}\underline{\alpha}
\wedge {{\partial L_{\bot}}\over{\partial\left({\it{l}}_{e_{_0}}
\underline{\alpha}\,\right)}} - L_{\bot}\,\biggr)\Biggr]\cr 
&\hskip1.1cm +\delta\alpha _{\bot}\wedge
\Biggl[ {{\partial L_{\bot}}\over{\partial\alpha _{\bot}}}
+(-1\,)^p\,{1\over{u^0}}\,\underline{d}\,\biggl( u^0\,
{{\partial L_{\bot}}\over{\partial\bigl(\underline{d}\,
\alpha _{\bot}\bigr)}}\biggr)\Biggr]\cr 
&\hskip1.1cm +\delta\underline{\alpha }\,\,\wedge 
\Biggl[ {{\partial L_{\bot}}\over{\partial\underline{\alpha}}}
-(-1\,)^p\,{1\over{u^0}}\,\underline{d}\,\biggl( u^0\,
{{\partial L_{\bot}}\over{\partial\bigl(\underline{d}\,
\underline{\alpha}\bigr)}}\biggr) 
-{\it{l}}_{e_{_0}}\Biggl( {{\partial L_{\bot}}\over{\partial 
\left({\it{l}}_{e_{_0}}\underline{\alpha}\,\right)}}\Biggr)
\,\Biggr]\Biggr\}\cr  
&+d\,\Biggl\{\delta\underline{\alpha }\wedge {{\partial L_{\bot}}
\over{\partial\left({\it{l}}_{e_{_0}}\underline{\alpha}\,\right)}}
-\vartheta ^0\wedge\Biggl[ \delta u^0\,{{\partial L_{\bot}}
\over{\partial\bigl(\underline{d}\,u^0\bigr)}}
+\delta\alpha _{\bot}\wedge {{\partial L_{\bot}}
\over{\partial\bigl(\underline{d}\,
\alpha _{\bot}\bigr)}} +\delta\underline{\alpha }\wedge 
{{\partial L_{\bot}}\over{\partial\bigl(
\underline{d}\,\underline{\alpha}\bigr)}}\Biggr]\Biggr\}
\cr }\eqno(C.3)$$
gives rise to the field equations
$$\eqalign{ {{\partial L_{\bot}}\over{\partial u^0}}
&-{1\over{u^0}}\,\underline{d}\,\biggl( u^0\,{{\partial L_{\bot}}
\over{\partial\bigl(\underline{d}\,u^0\bigr)}}\biggr)
-{1\over{u^0}}\,\biggl( {\it{l}}_{e_{_0}}\underline{\alpha}
\wedge {{\partial L_{\bot}}\over{\partial\left({\it{l}}_{e_{_0}}
\underline{\alpha}\,\right)}} - L_{\bot}\,\biggr)=\,0\cr 
{{\partial L_{\bot}}\over{\partial\alpha _{\bot}}}
&+(-1\,)^p\,{1\over{u^0}}\,\underline{d}\,\biggl( u^0\,
{{\partial L_{\bot}}\over{\partial\bigl(\underline{d}\,
\alpha _{\bot}\bigr)}}\biggr)=\,0\cr 
{{\partial L_{\bot}}\over{\partial\underline{\alpha}}}
&-(-1\,)^p\,{1\over{u^0}}\,\underline{d}\,\biggl( u^0\,
{{\partial L_{\bot}}\over{\partial\bigl(\underline{d}\,
\underline{\alpha}\bigr)}}\biggr) 
-{\it{l}}_{e_{_0}}\Biggl( {{\partial L_{\bot}}\over{\partial 
\left({\it{l}}_{e_{_0}}\underline{\alpha}\,\right)}}\Biggr)
=\,0\,.\cr }\eqno(C.4)$$
From (C.2) we define the only nonvanishing momentum (remember that
we will not study the constraints at this stage)
$$^\#\pi :={{\partial L_{\bot}}\over{\partial 
\left({\it{l}}_{e_{_0}}\underline{\alpha}\,\right)}}\,,\eqno(C.5)$$
and we define the Hamiltonian
$${\cal{H}}:=\,\,u^0\left(\,{\it{l}}_{e_{_0}}
\underline{\alpha}\wedge {}^\#\pi -L_{\bot}\right)\,.\eqno(C.6)$$
The variation of (C.6) reads 
$$\eqalign{\delta {\cal{H}} =&\,u^0\,\Biggl\{ 
\,\,{\it{l}}_{e_{_0}}\underline{\alpha}
\wedge \delta\,{}^\#\pi \cr 
&\hskip0.6cm -\delta u^0\,\Biggl[ {{\partial L_{\bot}}\over
{\partial u^0}}-{1\over{u^0}}\,\underline{d}\,\biggl( u^0\,
{{\partial L_{\bot}}\over{\partial\bigl(\underline{d}\,u^0
\bigr)}}\biggr)
-{1\over{u^0}}\,\biggl(\,{\it{l}}_{e_{_0}}
\underline{\alpha}\wedge\,{}^\#\pi -L_{\bot}\,\biggr)\Biggr]\cr 
&\hskip0.6cm -\delta\alpha _{\bot}\wedge
\Biggl[ {{\partial L_{\bot}}\over{\partial\alpha _{\bot}}}
+(-1\,)^p\,{1\over{u^0}}\,\underline{d}\,\biggl( u^0\,
{{\partial L_{\bot}}\over{\partial\bigl(\underline{d}\,
\alpha _{\bot}\bigr)}}\biggr)\Biggr]\cr 
&\hskip0.6cm -\delta\underline{\alpha }\,\,\wedge 
\Biggl[ {{\partial L_{\bot}}\over{\partial\underline{\alpha}}}
-(-1\,)^p\,{1\over{u^0}}\,\underline{d}\,\biggl( u^0\,
{{\partial L_{\bot}}\over{\partial\bigl(\underline{d}\,
\underline{\alpha}\bigr)}}\biggr)\,\Biggr]\,\Biggr\}\cr 
&-\underline{d}\,\Biggl\{\,u^0\,\Biggl[ \delta u^0\,
{{\partial L_{\bot}}\over{\partial\bigl(\underline{d}
\,u^0\bigr)}} +\delta\alpha _{\bot}\wedge {{\partial 
L_{\bot}}\over{\partial\bigl(\underline{d}\,
\alpha _{\bot}\bigr)}} +\delta\underline{\alpha }\wedge 
{{\partial L_{\bot}}\over{\partial\bigl(
\underline{d}\,\underline{\alpha}\bigr)}}\Biggr]\Biggr\}
\,.\cr }\eqno(C.7)$$
Making use of the field equations (C.4) and neglecting the total
divergence contributions, (C.7) transforms into
$$\delta {\cal{H}} =\,u^0\left[\,{\it{l}}_{e_{_0}}
\underline{\alpha}\wedge \delta\,{}^\#\pi 
-\delta\underline{\alpha }\wedge{\it{l}}_{e_{_0}}
\Biggl( {{\partial L_{\bot}}\over{\partial\left(
{\it{l}}_{e_{_0}}\underline{\alpha}\,\right)}}
\Biggr)\,\right] =\,u^0\left[\,{\it{l}}_{e_{_0}}
\underline{\alpha}\wedge \delta\,{}^\#\pi 
-\delta\underline{\alpha }\wedge {\it{l}}_{e_{_0}}
{}^\#\pi\,\,\right]\,.\eqno(C.8)$$
On the other hand, being 
$${\cal{H}} =\,{\cal{H}}\left(\,u^0\,,\,\alpha _{\bot}
\,,\,\underline{\alpha}\,,\,^\#\pi\,\right)\,,\eqno(C.9)$$
we find alternatively
$$\delta {\cal{H}}=\,\delta u^0\,
{{\delta {\cal{H}}}\over{\delta u^0}}
+\delta\alpha _{\bot}\wedge 
{{\delta {\cal{H}}}\over{\delta\alpha _{\bot}}}
+\delta\underline{\alpha}\wedge 
{{\delta {\cal{H}}}\over{\delta\underline{\alpha}}}
+{{\delta {\cal{H}}}\over{\delta\,{}^\#\pi}}
\wedge \delta\,{}^\#\pi \,.\eqno(C.10)$$ 
Comparing (C.8) with (C.10), the Hamilton equations follow, namely
$${{\delta {\cal{H}}}\over{\delta u^0}}=\,0
\quad\,,\qquad 
{{\delta {\cal{H}}}\over{\delta\alpha _{\bot}}}=\,0
\,,\eqno(C.11)$$
and 
$$u^0\,{\it{l}}_{e_{_0}}\underline{\alpha}=\,
{{\delta {\cal{H}}}\over{\delta\,{}^\#\pi}}
\quad\,,\qquad
u^0\,{\it{l}}_{e_{_0}}{}^\#\pi 
=-{{\delta {\cal{H}}}\over{\delta\underline{\alpha}}}
\,.\eqno(C.12)$$ 
Eqs.(C.11) have to do with the existence of constraints and are
treated more rigorously in the text. The important thing we
wanted to demonstrate here is that (C.12) coincide with (4.8) as
deduced shortly in section 4.
\bigskip\bigskip

\noindent {\bf D.--Poisson brackets}
\bigskip
We denote the canonically conjugated variables in a compact form 
as $Q^a_i\,,{}^\#\Pi _b^i\,$. Let $\omega $ be a p--form and 
$\sigma $ a q--form depending on these variables. We generalize
the definition (4.10) of the Poisson brackets in the obvious form 
$$\eqalign{\left\{\omega\,,\sigma \,\right\} &:=
{{\delta \sigma }\over{\delta\,{}^\#\Pi _a^i}}
\wedge {{\delta \omega}\over{\delta Q^a_i}}
-{{\delta \omega}\over{\delta\,{}^\#\Pi _a^i}}
\wedge {{\delta \sigma }\over{\delta Q^a_i}}\cr 
&\hskip0.1cm =\,\left\{ Q^a_i\,,\sigma\,\right\}
\wedge {{\delta \omega}\over{\delta Q^a_i}}
+{{\delta \omega}\over{\delta\,{}^\#\Pi _a^i}}
\wedge \left\{ {}^\#\Pi _a^i\,,\sigma\,\right\}\,.\cr }
\eqno(D.1)$$
It follows
$$\left\{\omega\,,\sigma \,\right\} 
=-\left\{\sigma \,,\omega\,\right\}\,,\eqno(D.2)$$
$$\left\{\omega\wedge\alpha\,,\sigma \,\right\} 
=\,\left\{\omega\,,\sigma \,\right\}\wedge\alpha  
+(-1)^{p\,(q+1)}\,\omega\wedge\left\{\alpha\,,\sigma \,\right\}
\,.\eqno(D.3)$$
In particular, the fundamental Poisson brackets satisfy 
$$\eqalign{\left\{ Q^a_i\,,Q^b_j\,\right\} =&\,0 \cr 
\left\{ {}^\#\Pi _a^i\,,{}^\#\Pi _b^j\,\right\} =&\,0 \cr 
\left\{ Q^a_i(x)\,,{}^\#\Pi _b^j(y\,)\,\right\} 
=&\,\delta ^a_b\,\delta _i^j\,\delta (x-y\,)\,,\cr }\eqno(D.4)$$
and in addition 
$$\left\{ \underline{D}\,Q^a_i\,,{}^\#\Pi _b^j\,\right\} 
=\,\underline{D}\,\left\{ Q^a_i\,,{}^\#\Pi _b^j\,\right\} 
\,.\eqno(D.5)$$
Being $Q^a_i$ not a vector but a connection $\underline{A}^a\,$,
we find 
$$\left\{ \underline{F}^a\,,{}^\#\Pi ^{\underline{A}} _b\,\right\} 
=\,\left\{ \underline{A}^a
\,,\underline{D}{}^\#\Pi ^{\underline{A}} _b\,\right\} 
=\,\underline{D}\,\left\{ A^a\,,{}^\#\Pi ^{\underline{A}}_b\,\right\} 
\,.\eqno(D.6)$$
On the other hand, being $\varphi ^{(1)}_a\,$, see (6.19b), a
first class constraint, it is the generator of a symmetry,
namely of the rotations. In fact, for any vector--valued form 
$\omega ^a$ holds 
$$\left\{\,\omega ^a\,,\varphi ^{(1)}_b\,\right\} 
=\,\epsilon ^a{}_{bc}\,\omega ^c\,,\eqno(D.7)$$
so that 
$$\delta\,\omega ^a 
=\,\lambda ^b\,\left\{\,\omega ^a\,,\varphi ^{(1)}_b\,\right\} 
=\,\epsilon ^a{}_{bc}\,\lambda ^b\,\omega ^c\,,\eqno(D.8)$$
whereas for the $SO(3)$ connection we find
$$\delta\,\underline{A}^a 
=\,\lambda ^b\,\left\{\,\underline{A}^a\,,\varphi ^{(1)}_b\,\right\} 
=-\underline{D}\,\lambda ^b\,.\eqno(D.9)$$
Thus, from the relation 
$$\delta\,\underline{D}\,\omega ^a 
=\,\underline{D}\,\delta\omega ^a 
+\epsilon ^a{}_{bc}\,\delta A^b\wedge\omega ^c\,,\eqno(D.10)$$
we find 
$$\left\{\,\underline{D}\,\omega ^a\,,\varphi ^{(1)}_b\,\right\} 
=\,\underline{D}\,\left\{\,\omega ^a\,,\varphi ^{(1)}_b\,\right\}
\,.\eqno(D.11)$$
This and similar properties were used in the calculation of the
constraints and of their stability conditions.
\bigskip\bigskip 

\noindent{\bf{E.--Several useful formulae}}
\bigskip 
The Hodge dual star in three dimensions is defined as follows 
$$\eqalign{^{\#}\left(\vartheta _a\wedge\vartheta _b
\wedge\vartheta _c\,\right)
&:={\overline{\eta }}_{abc}:=\,\eta _{0abc}=-e_{_0}\rfloor
\eta _{abc}\cr 
^{\#}\left(\vartheta _a\wedge\vartheta _b\,\right)
&:=\,{\overline{\eta }}_{ab} 
:=\,{\overline{\eta }}_{abc}\vartheta ^c
=\,e_{_0}\rfloor\eta_{ab}\cr  
^{\#}\vartheta _a &:=\,{\overline{\eta }}_a
:=\,{1\over{2!}}\,{\overline{\eta }}_{abc}
\vartheta ^b\wedge\vartheta ^c =-e_{_0}\rfloor\eta_a\cr  
^{\#}1 &:=\,{\overline{\eta }}
:=\,{1\over{3!}}\,{\overline{\eta }}_{abc}
\vartheta ^a\wedge\vartheta ^b\wedge\vartheta ^c 
=\,e_{_0}\rfloor\eta\,.\cr}\eqno(E.1)$$ 
The variation of dual forms reads, see Ref.(13),
$$\delta \,{}^*\alpha =\,{}^*\delta\alpha 
-{}^*\left(\delta \vartheta ^\alpha\wedge e_\alpha\rfloor\alpha\,\right) 
+\delta\vartheta ^\alpha\wedge\left( 
e_\alpha\rfloor {}^*\alpha\,\right)\,.\eqno(E.2)$$
Making use of the relations
$$^{\#}\alpha _{\bot}
=-\,{}^*\left(\vartheta ^0 \wedge\alpha
_{\bot}\,\right)\,,\qquad ^{\#}\underline{\alpha}\,\,
=-\,{}^*\left(\vartheta ^0\wedge\alpha\,\right)\,,\eqno(E.3)$$
we find
$$\eqalign{\delta \,{}^{\#}\alpha _{\bot} 
&=\,{}^{\#}\delta\alpha _{\bot} 
-{}^{\#}\left(\delta \vartheta ^a
\wedge e_a\rfloor\alpha _{\bot}\,\right) 
+\delta\vartheta ^a\wedge\left( 
e_a\rfloor {}^{\#}\alpha _{\bot}\,\right)\cr 
\delta \,{}^{\#}\underline{\alpha}\,\,\, 
&=\,{}^{\#}\delta\underline{\alpha}\,\,\,
-{}^{\#}\left(\delta \vartheta ^a
\wedge e_a\rfloor\,\underline{\alpha}\,\,\right)\,\, 
+\delta\vartheta ^a\wedge\left( 
e_a\rfloor {}^{\#}\underline{\alpha}\,\,\right)\,.\cr }\eqno(E.4)$$
Other formulae necessary to reproduce the calculations of this
paper are the following$^{(6,7,12)}$ 
$$^{\#\#}\alpha =\,\alpha\,,\eqno(E.5)$$
$$^\#\left(\alpha\wedge\vartheta _\alpha\,\right) 
=\,e_\alpha\rfloor {}^\#\alpha\,.\eqno(E.6)$$
Being $\alpha$ and $\beta$ p--forms
$$^\#\alpha\wedge\beta =\,^\#\beta\wedge\alpha\,,\eqno(E.7)$$
$$\vartheta ^a\wedge e_a\rfloor\alpha =\,p\,\alpha\,.\eqno(E.8)$$
\bigskip\bigskip

\centerline{REFERENCES}\vskip1.0cm

\noindent\item{[1]} R. Utiyama,  {\it Phys. Rev.} {\bf 101} (1956) 1597

\item\qquad T. W. B. Kibble, {\it J. Math. Phys.} {\bf 2} (1961) 212

\item\qquad D. W. Sciama, {\it Rev. Mod. Phys.} {\bf 36} (1964) 463 and 1103

\item\qquad A. Trautman, in {\it Differential Geometry},
Symposia Mathematica Vol. 12 (Academic Press, London, 1973), p. 139

\item\qquad A. G. Agnese and P. Calvini, {\it Phys. Rev.} {\bf
D 12} (1975) 3800 and 3804 

\item\qquad P. von der Heyde, {\it Phys. Lett.} {\bf 58
A} (1976) 141

\item\qquad E.A. Ivanov and J. Niederle, {\it Phys. Rev.} {\bf D25} 
(1982) 976 and 988

\item\qquad D. Ivanenko and G.A. Sardanashvily, {\it
Phys. Rep.} {\bf 94} (1983) 1

\item\qquad E. A. Lord, {\it J. Math. Phys.} {\bf 27} (1986) 2415 and 3051

\noindent\item{[2]} A.B. Borisov and I.V. Polubarinov, {\it Zh.
ksp. Theor. Fiz.} {\bf 48} (1965) 1625, and V. Ogievetsky and 
I. Polubarinov, {\it Ann. Phys.} (NY) {\bf 35} (1965) 167

\item\qquad A.B. Borisov and V.I. Ogievetskii, {\it Theor.
Mat. Fiz.} {\bf 21} (1974) 329

\item\qquad A. Salam and J. Strathdee, {\it Phys. Rev.}
{\bf 184} (1969) 1750 and 1760

\item\qquad C.J. Isham, A. Salam and J. Strathdee, {\it Ann. of Phys.} 
{\bf 62} (1971) 98

\item\qquad L.N. Chang and F. Mansouri, {\it Phys.
Lett.} {\bf 78 B} (1979) 274, and {\it Phys. Rev.} {\bf D 17} (1978) 3168

\item\qquad K.S. Stelle and P.C. West, {\it Phys. Rev.} 
{\bf D 21} (1980) 1466

\item\qquad E.A. Lord, {\it{Gen. Rel. Grav.}} {\bf 19} (1987) 983, and 
{\it J. Math. Phys.} {\bf 29} (1988) 258

\noindent\item{[3]} A. L\'opez--Pinto, A. Tiemblo and R.
Tresguerres, {\it Class. Quantum Grav.} {\bf 12} (1995) 1327 

\noindent\item{[4]} S. Coleman, J. Wess and B. Zumino, {\it Phys. Rev.}
{\bf 117} (1969) 2239 

\item\qquad C.G. Callan, S. Coleman, J. Wess and B. Zumino, {\it Phys. Rev.}
{\bf 117} (1969) 2247

\item\qquad S. Coleman, {\it Aspects of Symmetry}. Cambridge
University Press, Cambridge (1985)

\noindent\item{[5]} K. Hayashi and T. Nakano, {\it Prog. Theor. Phys} 
{\bf 38} (1967) 491

\item\qquad K. Hayashi and T. Shirafuji, {\it Prog. Theor. Phys} 
{\bf 64} (1980) 866 and {\bf 80} (1988) 711

\item\qquad G. Grignani and G. Nardelli, {\it Phys. Rev.} 
{\bf D 45} (1992) 2719

\item\qquad E. W. Mielke, J.D. McCrea, Y. Ne'eman and F.W. Hehl 
{\it Phys. Rev.} {\bf D 48} (1993) 673, and references therein

\item\qquad J. Julve, A. L\'opez--Pinto, A. Tiemblo and R.
Tresguerres, {\it Nonlinear gauge realizations of spacetime
symmetries including translations}, to appear in G.R.G. (1996)

\noindent\item{[6]} F. W. Hehl, J. D. McCrea,  E. W. Mielke, and
Y. Ne'eman {\it Found. Phys.} {\bf 19} (1989) 1075

\item\qquad R. D. Hecht and F. W. Hehl, {\it {Proc. 9th Italian
Conf. G.R. and Grav. Phys.}}, Capri (Napoli). R. Cianci 
et al.(eds.) (World Scientific, Singapore, 1991) p. 246  

\item\qquad F.W. Hehl, J.D. McCrea, E.W. Mielke, and Y. Ne'eman, {\it
Physics Reports} {\bf 258} (1995) 1

\noindent\item{[7]} F.W. Hehl, P. von der Heyde, G. D. Kerlick
and J. M. Nester, {\it Rev. Mod. Phys.} {\bf 48} (1976) 393

\item\qquad F.W. Hehl, {\it Proc. of the 6th Course of the School of
Cosmology and Gravitation on Spin, Torsion, Rotation, and
Supergravity}, held at Erice, Italy, May 1979, P.G. Bergmann, V.
de Sabbata, eds. (Plenum, N.Y. i980) 5

\noindent\item{[8]} P. Baekler, F.W. Hehl, and E.W. Mielke, 
{\it Proc. of the 2nd Marcel Grossmann Meeting on Recent
Progress of the Fundamentals of General Relativity 1978}, R.
Ruffini, ed. (North--Holland, Amsterdam 1982) 413

\item\qquad P. Baekler, F.W. Hehl, and H.J. Lenzen, {\it Proc. 
of the 3rd Marcel Grossmann Meeting on General Relativity}, Hu
Ning, ed. (North--Holland, Amsterdam 1983) 107

\item\qquad J.D. McCrea, P. Baekler and M. G\"urses, {\it Nuovo 
Cimento} {\bf B 99} (1987) 171

\item\qquad P. Baekler, M. G\"urses, F.W. Hehl and J.D. McCrea, 
{\it Phys. Lett.} {\bf A 128} (1988) 245

\item\qquad R. Tresguerres, {Z. Phys.} {\bf C 65} (1995) 347,
and {\it Phys. Lett.} {\bf A 200} (1995) 405

\noindent\item{[9]} A. Ashtekar, {\it Phys. Rev. Lett.} {\bf 57}
(1986) 2244

\item\qquad A. Ashtekar, {\it Phys. Rev.} {\bf D 36} (1987) 1587

\item\qquad A. Ashtekar, {\it Non Perturbative Canonical Gravity}  
(Notes prepared in collaboration with R. S. Tate). (World 
Scientific Books, Singapore, 1991)

\noindent\item{[10]} P.A.M. Dirac, {\it Canadian Jour. of Math.}
{\bf 2} (1950) 129

P.A.M. Dirac, {\it Lectures on Quantum
Mechanics}, Belfer Graduate School of Science, Yeshiva
University, N.Y. (1964)

\noindent\item{[11]} A. Hanson, T. Regge and C. Teitelboim, {\it
Constrained Hamiltonian Systems}, Roma, Accademia dei Lincei
(1976) 

\noindent\item{[12]} E.W. Mielke, {\it Phys. Lett.} {\bf A 149}
(1990) 345 

\item\qquad E.W. Mielke, {\it Ann. Phys.} (N.Y.) {\bf 219}
(1992) 78 

\noindent\item{[13]} R.P. Wallner, Ph.D. Thesis, University of
Vienna (1982)

\item\qquad R.P. Wallner, {\it Phys. Rev.} {\bf D 42} (1990) 441

\item\qquad R.P. Wallner, {\it Jour. Math. Phys.} {\bf 36}
(1995) 6937

\noindent\item{[14]} E.P. Wigner, {\it Ann. Math.} {\bf 40}
(1939) 149 

\item\qquad S. Weinberg, {\it Phys. Rev.} {\bf 133 B}
(1964) 1318

\noindent\item{[15]} F.W. Warner, {\it Foundations of 
differentiable manifolds and Lie groups}, {\it Scott, Foresman 
and Company}, Glenview, Illinois (1971)

\noindent\item{[16]} J. F. Barbero G., {\it Physical Review} 
{\bf D 51} (1995) 5507

\bye